\documentclass[aps,pre,showpacs,floatfix,twocolumn,superscriptaddress,longbibliography]{revtex4-2}

\usepackage{mathtools,amssymb,lipsum, nccmath}
\usepackage[T1]{fontenc}
\usepackage{textcomp}
\usepackage{graphicx}
\usepackage{bm,amsmath,amssymb}
\usepackage{physics}
\usepackage{dcolumn}
\usepackage{subfigure}
\usepackage{color}
\usepackage{tikz}
\usepackage{fontawesome}
\usepackage{esint} 
\usepackage{ifdraft}
\usepackage{changes}
\usepackage{lipsum}% <- For dummy text
\definechangesauthor[name={TCS}, color=red]{TCS}

\newcommand\link[1]{\href{#1}{\faLink}}
\DeclareMathOperator{\La}{\mathcal{L}}

\usepackage[%  
    colorlinks=true,
    pdfborder={0 0 0},
    linkcolor=red,
    citecolor = red
]{hyperref}
\usepackage{ifthen}
\usepackage{mathpazo}

\makeatletter
\renewcommand{\maketag@@@}[1]{\hbox{\m@th\normalsize\normalfont#1}}%
\makeatother
\begin{document}
\title{Snake net and balloon force with a neural  network  for detecting multiple phases
}

\author{Xiaodong Sun}
\affiliation{College of Physics, Taiyuan University of Technology, Shanxi 030024, China} 

\author{Huijiong Yang}
\affiliation{College of data science, Taiyuan University of Technology, Shanxi 030024, China}
\affiliation{Department of computing, The Hong Kong Polytechnic University, Hong Kong, China} 

\author{Nan Wu}
\affiliation{College of Physics, Taiyuan University of Technology, Shanxi 030024, China} 

\author{T.C. Scott}
\affiliation{Institut f\"{u}r Physikalische Chemie, RWTH Aachen University, Aachen 52056, Germany}

\author{Jie Zhang}
\affiliation{College of Physics, Taiyuan University of Technology, Shanxi 030024, China}

\author{Wanzhou Zhang}
\email{corresponding author: zhangwanzhou@tyut.edu.cn}
\affiliation{College of Physics, Taiyuan University of Technology, Shanxi 030024, China} 
\affiliation{
CAS Key Laboratory for Theoretical Physics, Institute of Theoretical Physics, Chinese Academy of Sciences, Beijing 100190, China}

\affiliation{
Hefei National Laboratory for Physical Sciences at the Microscale and Department of Modern Physics, University of Science and Technology of China, Hefei 230026, China}

\date{\today}

\begin{abstract}
Unsupervised machine learning applied to the study of phase transitions is an
ongoing and interesting research direction. 
The active contour model, also called the snake model, was initially proposed  
 for target contour extraction in two-dimensional
images. In order to obtain  a physical phase diagram, the snake model with an artificial neural network
is applied in an unsupervised learning way by the authors of  [Phys.Rev.Lett. 120, 176401(2018)]. It guesses the phase boundary as an initial snake and then drives the snake to convergence with forces estimated by the artificial neural network.
In this paper, we extend this unsupervised learning  method with one contour to a 
snake net 
   with multiple contours for the purpose of obtaining several phase boundaries in a phase diagram.
For the classical Blume-Capel model, the phase diagram containing three and four phases is obtained. 
Moreover, to overcome the limitations of the initial position and speed up the movement of the snake, the balloon force decaying with the iteration steps is  introduced and applied to the 
snake net 
structure. Our method is helpful in determining the phase diagram with multiple phases, using just snapshots of configurations from cold atoms or other experiments without knowledge of the phases.

\end{abstract}

\maketitle

\section{introduction}
Exploring the phases and phase diagrams of 
 the matter is a long-standing task in physics ~\cite{review}. Commonly found in life, such as water, there are three phases: solid-liquid-gas. In addition, states of matter exist at very low 
temperatures such as superconductors~\cite{sc}, superfluids~\cite{sf}, and, at very high temperatures, plasma states~\cite{plasma}. The study of the distribution of these phases in the phase diagram and the phase transition boundaries between them is very helpful for one to understand the natural world.

\begin{figure*}[t]
  %  \flushleft
    \includegraphics[width=17.cm,height=4.6cm]{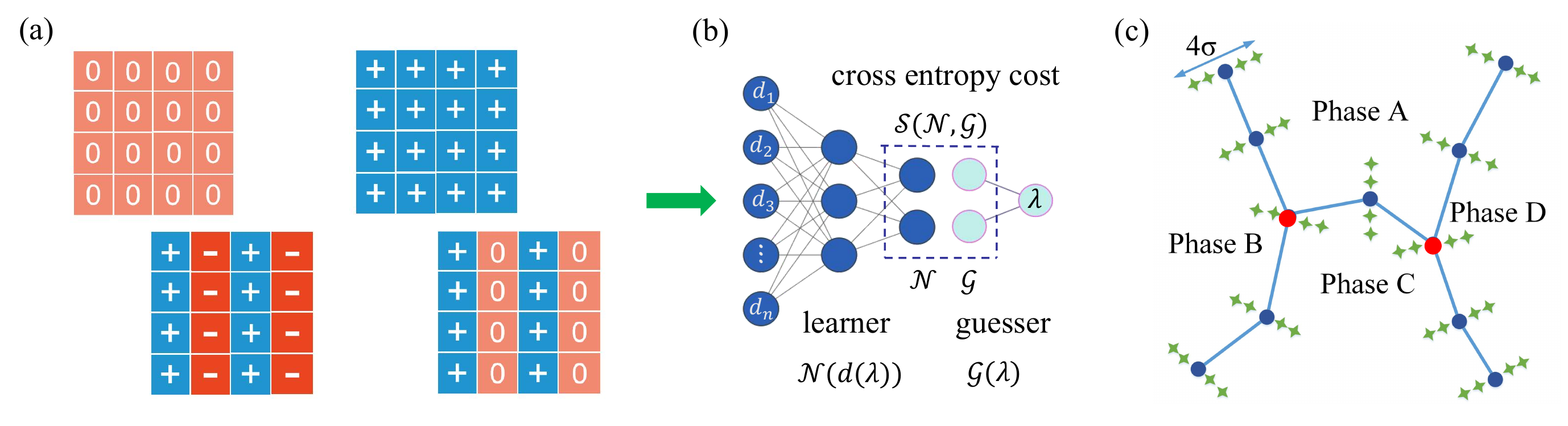}
       \caption{ The basic idea of using the SN-DCN to locate the phase boundaries. (a) The typical input  configurations of the BC model. (b) The DCN tool processes the data. The DCN  contains a learner network $\mathcal{N}$ and a guesser network $\mathcal{G}$. (c)  The phase boundaries detected by the SN-DCN. Blue circles mark the nodes of the snake model, and the red circle indicates a special kind of node where the three snakes intersect. The  unit width of the snake is indicated by $\sigma$ (green diamond) and its value is dynamic. The line segments between nodes indicate the phase boundaries to be detected.}
\label{fig:FIG1}
\end {figure*}
%One first obtains the dataset of the system, such as the configuration of the spin systems. Then one initializes the position of the NSS according to the necessary  prior knowledge and applies a DCN containing a guesser network $\mathcal{G}$ and a learner network $\mathcal{N}$ to drive the initial snakes. Eventually, the snakes locates to the phase boundary, \added[id=TCS]{as shown} in Fig.~\ref{fig:FIG1} (c).
%One first obtains the dataset of the system, such as the configuration of the spin systems. Then one initializes the position of the NSS according to the necessary  prior knowledge and applies a DCN containing a guesser network $\mathcal{G}$ and a learner network $\mathcal{N}$ to drive the initial snakes. Eventually, the snakes locates to the phase boundary, \added[id=TCS]{as shown} in Fig.~\ref{fig:FIG1} (c).

With the development of machine learning methods and their integration into various disciplines, machine learning methods are  used to study the phases of matter~\cite{review, Juan}. Unsupervised machine learning does not require real labels for the data and is therefore more appreciated by researchers when studying unknown questions.
Commonly used unsupervised machine learning methods are principal component analysis \cite{pca1,pca2,pca3},  t-distributed stochastic neighbor embedding~\cite{ PhysRevE.97.013306, PhysRevB.103.075106}, and diffusion maps~\cite{df1, df2, df3,PhysRevResearch.3.013074}. 
Recent work also proposed a quantum algorithm to achieve a quantum computational speedup of diffusion maps~\cite{PhysRevA.104.052410}.

 In 2018, a  simple snake model with a neural network was proposed  to search for the phase boundaries between the two phases in 
the two-dimensional parameter space~\cite{liuprl}.
The snake model, also known as the active contour model, was originally proposed by Kass {it et al.} in the 1980s for target contour extraction in two-dimensional images in the field of computer vision ~\cite{kass}. The method  is useful because it replaces image processing
with the active contour energy minimization problem.
The movement of the active contour  is  driven by the well-defined image force ${\bf F}_{img}$.

For detecting phase transitions rather than images processing, The authors of Ref.~\cite{liuprl} proposed the neural network  discriminative cooperative network (DCN), 
which consists of a learner network $\mathcal{N}$ and a guesser network $\mathcal{G}$ and the two networks work in cooperation with each other.
The DCN  replaces the image force
 ${\bf F}_{img}$
of the snake model for images  with the derivative of the cross entropy cost function $\mathcal{S}$ between the outputs of $\mathcal{G}$ and $\mathcal{N}$ with respect to the position $\lambda_g$ of snakes, i.e., 
  $-\frac{\partial \mathcal{S}}{\partial \lambda_g}$.
  Using the method developed, 
 the boundary between the superfluid and insulating phases can be  obtained~\cite{liuprl}.
However, the simple snake model, i.e., only one contour with a DCN  still encounters challenges in the study of phase diagrams, especially for more two distinct  phases~\cite{liuprl}.

Here, we propose to combine the snake net (SN) and the DCN together to find multiple boundaries between phases.
The topology-preserving SN was developed by M. Butenuth  in 2012  for image contour extraction ~\cite{Butenuth2012,DBLP:conf/bildmed/ButenuthJ07}.  The SN has multiple snakes connected by common nodes.
By updating the positions of the snakes, the SN model  can realize  the  contours of images containing multiple colors and  the images
can be cells, roads, and so on.
 In our SN-DCN method, the snakes in the SN are expected to  converge to the real phase  boundaries. 
 %====================\\
%The network-shaped snake (NSS) model has many sub-snakes connected to form a network. The topology of this network  is the same as that of the  phase boundaries or image contours. 
%It is natural to ask whether the boundaries of a physical phase diagram containing multiple phases can be found.
%However, for physical systems, it needs to be tested further on whether or not  such an NSS model with \added[id=TCS]{the} DCN  can handle a more complex phase diagram. 

%On one hand,  it is also not clear how to use \added[id=TCS]{the} DCN to drive several sub-snakes to the target boundaries, and the difference  between the segmentation of an image and locating the phase boundaries. 
On the other hand, if the initial position of the snake is far from the true boundary, in such a case, the initial snake converges very slowly and does not even get to the correct position because the snake does not feel enough force.
Therefore, we introduce the balloon force~\cite{COHEN1991211} to the snake model,  originally proposed by L. Cohen {\em et al.} to help locate contours over image processing.  The balloon force (BF)  can be a good solution to the problematic requirement that the initial snake must be set near the true phase transition boundary.  
The BF can also speed up the movement of the snake and help reduce the number of iterative steps.
 
The basic idea is shown in Fig.~\ref{fig:FIG1}.
One first obtains the dataset of the system, such as the configuration of the spin systems. Then one initializes the position of the SN according to the necessary  prior knowledge and applies a DCN containing $\mathcal{G}$ and $\mathcal{N}$ to drive the initial snakes. Eventually, the snake locates at the phase boundary, as shown in Fig.~\ref{fig:FIG1} (c).

This work builds a SN-DCN to obtain the phase boundary of the physical systems of interest. Here, there are two versions of the SN-DCN. The simplest SN has only three snakes with a common node to get three-phase boundaries. The other extended version of the SN, with five snakes, yields five phase boundaries between the phases. 
The additional balloon force introduced here can also be selectively added to  the snake net. We also test the  hyperparameters of simple snake models with the DCN and find that the balloon force acceleration is most effective.

%To advance the NSS model with \added[id=TCS]{a} DCN to get the boundaries of the more general phase diagram, we apply the method ~\cite{DBLP:conf/bildmed/ButenuthJ07, three_snakes}   to  the spin-1  Blume-Capel (BC) model~\cite{BCmodel}, which contains configurations such as $\begin{smallmatrix} +-\\+- \end{smallmatrix}$, $\begin{smallmatrix} +0\\+0 \end{smallmatrix}$ , $\begin{smallmatrix} ++\\++ \end{smallmatrix}$, and  $\begin{smallmatrix} 00\\00 \end{smallmatrix}$ and so on

The outline of this paper is as follows. In Sec.~\ref{sec:theory}, we present  the SN-DCN method. In Sec.~\ref{sec:SN-DCN-app},  the method is applied to 
the Blume-Capel (BC)  model with 3 phases and 4 phases, respectively.  Different topologies of the SN are also discussed. In Sec.~\ref{sec:balloon}, the  BF-SN with the DCN  are presented and applied to the phase diagram for the quantum  Bose-Hubbard (BH) model and the BC model.
The conclusion and discussion are 
presented in Sec.~\ref{sec:con}. In appendix \ref{sec:appendixA}, a detailed description of the snake model and the iteration matrix of our models are presented. 

\section{The SN-DCN method}
\label{sec:theory}
\subsection{Input data}

\begin{figure}[ht]
 \includegraphics[height=3.4cm,width=8.4cm]{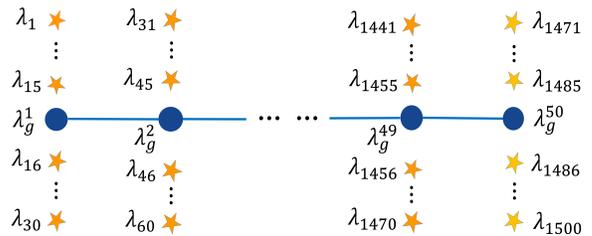}
 \caption{
 The way DCN gets input parameters. The blue circles represent the nodes  $\lambda^i_g$, $i=1,\cdots 50$. The orange pentagrams represent the sampled parameters collected.
 }  
 \label{fig:FIG2a}
\end{figure}

The input data type depends on the specific model and parameter range. For the BC model~\cite{BCmodel}, the input data are the spin configurations obtained from Metropolis Monte Carlo simulations~\cite{Metropolis1953EquationOS} as shown in Fig.~\ref{fig:FIG1} (a). 
The symbols ``+'', ``-'', and ``0'' correspond to the values taken by the spins. The stripe-like pattern corresponds to the states  $\begin{smallmatrix} +-\\+- \end{smallmatrix}$, or $\begin{smallmatrix} +0\\+0 \end{smallmatrix}$.
    The uniform pattern  corresponds to the states $\begin{smallmatrix} ++\\++ \end{smallmatrix}$, and $\begin{smallmatrix} 00\\00 \end{smallmatrix}$, respectively.
    The data comes from  a lattice with  the size  of $16\times 16$.
For the Bose-Hubbard models~\cite{BH}, the input data are wave functions in the mean-field framework and are expressed as the square of the expansion coefficients.
 
In real simulations,
the physical parameters can be temperature or different types of interaction  labeled by 
$\lambda$, which  usually has two components in the physical parameter space $(\lambda^x, \lambda^y)$.

As shown in Fig.~\ref{fig:FIG2a},
 each snake in the net has  50 nodes  marked by blue circles, i.e., $\lambda_g^i, i=1, 2\cdots 50$. They  also represent the position of the guessed phase transition points.
 The sampled parameters are denoted as  $\lambda_j, j=1, 2\cdots 1500$.
For each node, a line is drawn perpendicular to the snake with the node as the center point, and 30 sampling parameters are taken at uniform intervals on the line, whose
length is restricted to $[-2 \sigma, 2 \sigma]$ and $4 \sigma$ is 
also the width of the snakes. 
By simulating the BC model or the BH model with parameter $\lambda$,
one can get an (average) configuration or wave function $d(\lambda)$.
The input to $\mathcal{G}$ is the value of $\lambda$ and the input to $\mathcal{N}$ is
$d(\lambda)$  as shown in Fig.~\ref{fig:FIG1}.

\subsection{DCN}

Fig.~\ref{fig:FIG1} (b) shows the structure of the DCN, which  includes a learner network $\mathcal{N}$ and a guesser network $\mathcal{G}$.
The learner network  $\mathcal{N}$ is a fully-connected network that absorbs the classical configurations of the BC model  or wave functions of the BH model, labeled by $d(\lambda)$ and outputs their classifications $p_{A}^\mathcal{N}$ and $p_{B}^\mathcal{N}$, i.e.,  the probability that $d(\lambda)$ belongs to phases $A$ and $B$, respectively. The neurons in the hidden layer  reads $y_{H} = f(d(\lambda)\cdot{\bf W_{1}}+b_{1})$ and the neurons in the output layer  
yield
%ouput  
$\mathcal{N}(d(\lambda))=(p_{A}^\mathcal{N}, p_{B}^\mathcal{N})=f(y_{H}\cdot{\bf W_{2}} + {\bf b_{2}})$. Here  $\bf W_{1}$ and  $\bf W_{2}$ are  the weight matrices and $\bf b_{1}$ and $\bf b_{2}$ are the bias vectors.
The activation function for the neurons in the hidden and output layer is a sigmoid function.% defined as $f(x)=1/(1+e^{-x}$).

As shown in Fig.~\ref{fig:FIG1} (b), 
 the guesser $\mathcal{G}$ absorbs $\lambda$  and outputs two labels through the sigmoid function to determine the probability that $\lambda$ belongs to phase $A$ or $B$, defined as:
\begin{equation}
    \mathcal{G}_{A,B}(\lambda)=(p_{A}^\mathcal{G}, p_{B}^\mathcal{G})=sigmoid[s_{A,B}(\lambda-\lambda_{g})/\sigma],
\end{equation}
 where $s_{A,B}=-,+$. 
  The cross-entropy cost function between  $\mathcal{N}$ and $\mathcal{G}$ is defined as:
 \begin{equation}
   \mathcal {S}(\mathcal{N}, \mathcal{G})=-\mathcal{G} \cdot\log\mathcal{N}  - (1- \mathcal{G})\cdot \log(1-\mathcal{N}),
  \label{eq:equation2}
 \end{equation}
 and the smallest $\mathcal {S}$ indicates the best match between the guessed boundary and the true boundary. 
 
 %In the real simulation, $\lambda_{g}$ and $\sigma$ can perform gradient descent simultaneously.  
Similarly to the theory of generative adversarial networks~\cite{goodfellow2016deep}, the DCN simultaneously optimizes $\mathcal{N}$ and $\mathcal{G}$ to obtain the minimum  $\mathcal {S}$. $\mathcal{N}$  gets better learning results by updating the parameters ${\bf W}_{\mathcal{N}}$ ($\bf W_{1}$, $\bf W_{2}$, $\bf b_{1}$ and $\bf b_{2}$ ), and $\mathcal{G}$ gets better guessing results by updating $\lambda_{g}$ and $\sigma$. These parameters cooperate to achieve the purpose of discriminating between the two phases. The dynamics of both networks can be defined as:
\begin{subequations}
	\begin{align}
 \Delta {\bf W}_{\mathcal{N}}=-\alpha_{\mathcal{N}}\partial \mathcal {S}/\partial {\bf W}_{\mathcal{N}},\\
  \Delta\lambda_{g}=-\alpha_{\lambda_{g}}\partial \mathcal {S}/\partial \lambda_{g},\\
  \Delta\sigma=-\alpha_{\sigma}\partial \mathcal {S}/\partial \sigma,
	\end{align}
 \label{eq:dyn}
\end{subequations} 
where 
 $\alpha_{\mathcal{N}}$, $\alpha_{\lambda_{g}}$ and $\alpha_{\sigma}$ are the learning rates. 
 The partial derivatives of the above equations are expressed as:
\begin{subequations}
	\begin{align}
  \pdv{\mathcal {S}}{\vb{\mathcal{G}}}& = -\log\mathcal{N}+\log(1-\mathcal{N}),\\
  \frac{\partial \mathcal{G}_{A,B}}{\partial \lambda_{g}}& =-\frac{s_{A,B}}{4\sigma \cosh^2 [(\lambda-\lambda_{g})/2\sigma]},\\
  \frac{\partial \mathcal {G}}{\partial\sigma}& = \frac{\lambda-\lambda_{g}}{\sigma} \frac{\partial \mathcal{G}}{\partial \lambda_{g}}.
	\end{align}
\end{subequations}.

\begin{figure}[htb]
 \includegraphics[height=5cm,width=8.5cm]{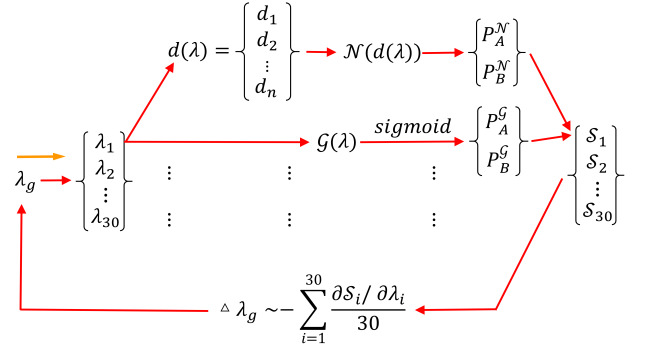}
 \caption{
 The process of updating the nodes ($\lambda_g$)  in the DCN. The orange arrow is at the beginning of the iterations.}  
 \label{fig:FIG2b}
\end{figure}

Fig.~\ref{fig:FIG2b} gives the flowchart of the update of the position of one node.
Starting from an initial input $\lambda_g$, then  one gets  the parameters labeled as  $\lambda_1, \cdots, \lambda_{30}$. For each  $\lambda_i$, there is  a $\mathcal{S}_i$ obtained by evaluation of cross-entropy. By averaging the 30 cross-entropies, $\Delta \lambda_g$ can be obtained and 
used to update $\lambda_g$ for the next round of iterations. 

For detecting the phase boundaries of  multiple phases, many nodes are usually required.
Our aim is to initialize the nodes in the parameter plane and drive them all close to the real phase transition boundary using an active contour method or snake model introduced in the next section.

\subsection{The SN model}

%$\lambda-\lambda_g$ is sampled uniformly between $-2\sigma$  and $2\sigma$, illustrated by the green symbols in Fig.~\ref{fig:FIG1} (c).

% and the selected sample is perpendicular to the direction of the node of the snake.
%For each node, the force on it is the average force on sample points   i.e., the green symbols around the node.

\subsubsection{The simple snake model}
The snake model is defined as a parametric contour,
\begin{equation}
     	C(s,t)=(x(s,t),y(s,t)),
\end{equation}
where $s\in [0,1]$ is a parameter, and $C(0,t)=C(1,t)$ for a closed contour  where the boundary is periodic. $t$ is  the number 
of iterations.  For the  images,  $x(y)$ refers to the real position of the nodes. For the  physical phase diagram to be studied, $x(y)$ represents the value of physical parameters such as temperature, or the interactions.

The total energy $E$ is composed of the internal energy $E_{int}$ and the external energy $E_{ext}$ or the image energy $E_{img}$.  
The snake in the image has total energy given by:
\begin{equation}
E(C)=\int_{0}^{1}[E_{int}(C(s))+E_{ext}(C(s))]ds,
  \label{eq:total_energy}
\end{equation}
where the internal energy reads:
\begin{equation}
	E_{int}=\frac{1}{2}\left[\alpha(s)| C^{'}(s)|^2+\beta(s)|C^{''}(s)|^2\right].
   \label{eq:equation7}
\end{equation}
In the equation above, $C^{'}(s)$ and $C^{''}(s)$ are the first derivative  and the second derivative  of $C(s)$ with respect to $s$.  The parameters  $\alpha(s)$ and $\beta(s)$ are adjustable and   control the 
continuity and smoothness of the curve. 
%{\textcolor{red}{$E_{ext}(C(s))=E_{img}(C(s))+E_{con}(C(s))$, since $E_{con}$ is a term that can be omitted in our model, to simplify the description, $E_{img}$ as it appears in this article is equivalent to $E_{ext}$.}}
Here, the external energy is restricted to  the image energy:
\begin{equation}
    E_{img}(C(s))=-|\partial{G_\sigma(C)[I(C)}]|^2,
    \label{Eimg}
\end{equation} where
 $I(C)$ is the value of pixels, $\partial$ is a gradient 
operator, and $G_{\sigma}$ is a two-dimensional Gaussian Kernel. The process of minimizing the total energy of the snake will allow the position of the snake to coincide with the boundary of the target object.
 The snake is driven by the image force:
 \begin{equation}
     {\bf F}_{x(y)}= -\frac{\partial E_{img} }{\partial x(y) }.
     \label{eq:equation4}
 \end{equation}
In the framework of the DCN, $E_{img}$ is replaced by  $\mathcal {S}$, and the node coordinates $(x,y)$ are replaced with the physical parameter 
   $(\lambda_g^x, \lambda_g^y)$.

%{\textcolor{red}{The phases $A$, $B$, $C$, and $D$ are surrounded by five sub snakes which are separated by red circles.}

\subsubsection{The snake net model}

\begin{figure}[ht]
 \includegraphics[height=4.6cm,width=7.5cm]{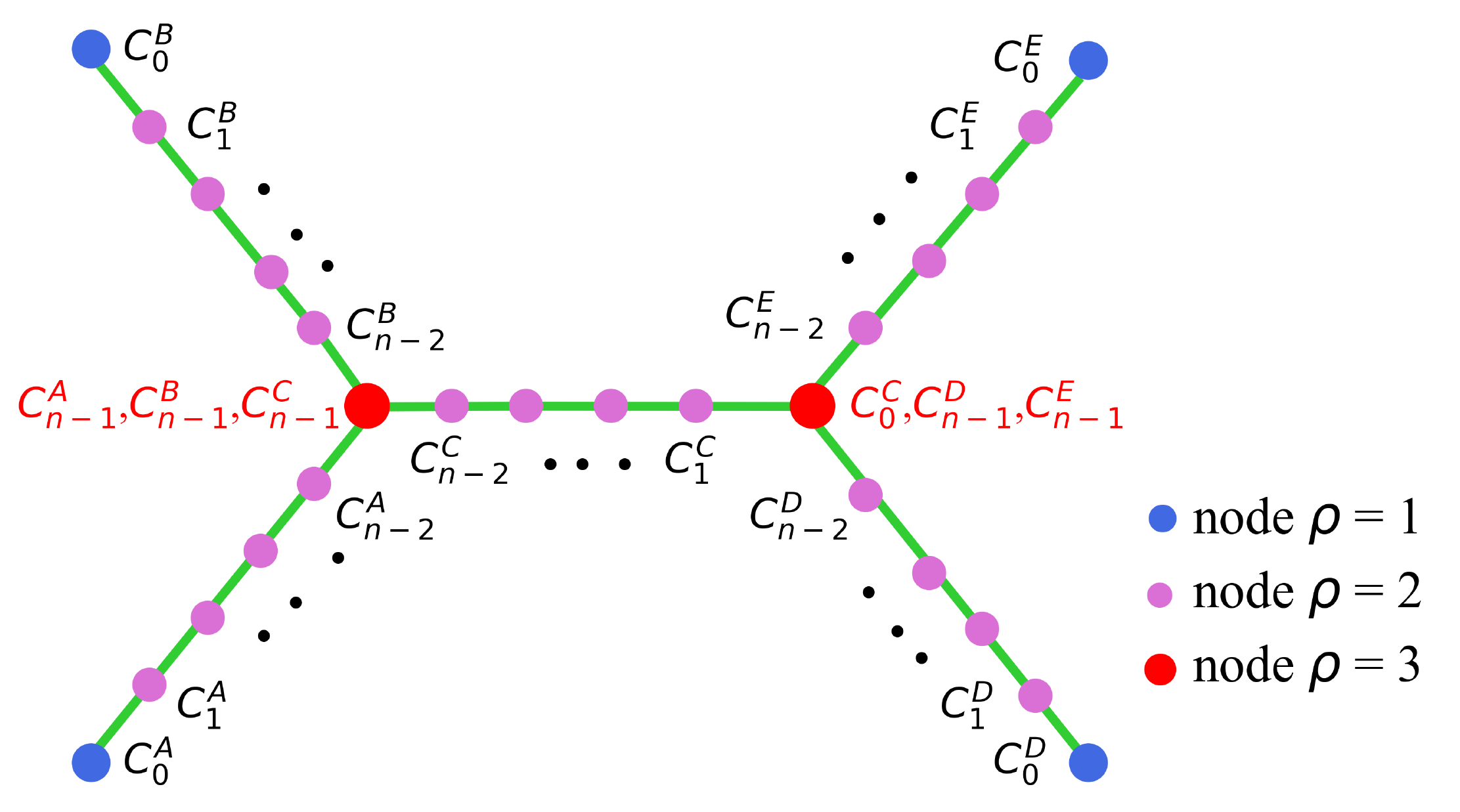}
 \caption{
 Topology of the extended  SN model. The SN model contain five snakes $C^{A}_{i} - C^{E}_{i}$. These five snakes have three different nodes $\rho=1$, $\rho=2$, and $\rho=3$, respectively.}
 \label{fig:FIG2}
\end{figure}

In Fig.~\ref{fig:FIG2}, the five snakes are separated by red circles. The snakes  are denoted by $C_i^j$, where $j=A, B, C, D, E$ means five snakes and $i=0,\cdots, n-1$ denotes the nodes of the snakes. In total, there are three kinds of nodes, 
  characterized by their degrees  $\rho(C)$. Specifically,
 $\rho(C) = 1, 2$ denote the outer endpoints and the inner nodes, respectively, which can be driven by the force similar to those in simple snakes.    $\rho(C) = 3$ denotes the common node of the different snakes~\cite{Butenuth2012}. 

For a pure image, by the minimization of the energy functional 
Eq.~\eqref{eq:total_energy},
the different kinds of nodes satisfy the following  differential equation:
\begin{equation}
   - \alpha \vb{C}'' + \beta \vb{C}'''' +\frac{-\partial|\partial{G_\sigma(C)[I(C)}]|^2}{\partial C} =0,
   \label{eq:one1}
\end{equation}
whose detailed description is given in appendix~\ref{sec:appendixA}.
The above equation 
gives the best description of effects of internal forces \({\bf F}_{int}= \alpha \vb{C}'' - \beta \vb{C}''''\) and external forces \({\bf F}_{ext}=-\grad E_{img}\) on each node of a snake.
Switching from the pure images to the physical systems, the external forces have to be replaced by $\Delta\lambda_{g}$.

For a closed, i.e., periodic boundary condition, a finite difference operation on the Eq.~(\ref{eq:one1}) yields:
\begin{align}
     	&\frac{\partial E_{img}}{\partial C_i}+\alpha((C_{i}-C_{i-1})-(C_{i+1}-C_{i}))\nonumber\\
     	+&\beta(C_{i-2}-2C_{i-1}+C_{i})-2\beta(C_{i-1}-2C_{i}+C_{i+1})\nonumber\\
     	+&\beta(C_{i}-2C_{i+1}+C_{i+2})=0.
\label{eq:EQ8}
\end{align}
However, for the SN model, each snake has a common node $C_{n-1}$  marked in red, which obeys:
\begin{align}
     	&\frac{\partial E_{img}}{\partial C_{n-1}}+\xi[(C_{n-1}-C_{n-2})-(C_{n-2}-C_{n-3})]=0, 
\label{eq:EQ9}
\end{align}
where only the first term of Eq.~(\ref{eq:EQ8}) with $\beta$ as a coefficient, is retained and the other terms, for example, $C_{n}$ and $C_{n+1}$, are not present at the ending points.  Here $\xi$ is another parameter to be controlled~\cite{Butenuth2012}.

%{\textcolor{blue}{we omit the relevant phases since $C_{n}^j$ and $C_{n+1}^j$ do not exist according to Eqs.~(\ref{eq:EQ8}), and $\xi$ is another parameter to be controlled~\cite{Butenuth2012}.}}

By combining the set of equations for all nodes together  (see Appendix~\ref{sec:appendixA2}), the following iterative equation can be obtained,
     \begin{equation}
     	{\bf A}C+\eta f(C)=0,
     	\label{eq:EQ10}
     \end{equation} 
where ${\bf A}$ is a pentadiagonal band matrix, which only depends on the parameters $\alpha$, $\beta$, and $\xi$. $\eta f(C)  = \frac{\partial E_{img}}{\partial C}$, where $\eta$ is an additional parameter 
to control the weight between internal and external energy. The iteration steps for the snakes between  $C_{next}$ and  $C_{current}$ are:
\begin{equation}
     	C_{next}=({\bf A}+\gamma {\bf I})^{-1}[\gamma C_{current}-\eta f(C_{current})]
     	\label{eq:ct},
     \end{equation}  
where $I$ is the identity matrix and $\gamma$ is the step size of the snakes.

Fig.~\ref{fig:FIG2} only shows
the extended SN. Sometimes, a simple SN can be used with three snakes. 
The difference between  the simple SN and the extended SN is the number of common nodes.
The former has only one common node and the latter has more than one common node.
Moreover, 
mathematically, the iteration of Eq.~(\ref{eq:ct}) can be different.
In the appendix \ref{sec:appendixA2}, three types of matrices $\bf {A}$ are shown for  (i) a closed snake, (ii) a snake with a fixed node at one end, and a common node at the other end. (iii) a snake with both endpoints as common nodes.

For image segmentation~\cite {Butenuth2012}, a big matrix ${\bf A}$ can contain the elements for all snakes. The couplings between different snakes are defined in ${\bf A}$. For physical systems, we separate the big matrix into several small matrices for each snake, and the coupling between them is realized by passing the positions of the common nodes.  %The matrix ${\bf A}$ is discussed  in the appendix~\ref{sec:appendixA2}.

\section{The application of the SN-DCN method}
\label{sec:SN-DCN-app}

\subsection{The BC model}

We choose the BC model~\cite{BCmodel} to test our method. The BC model  on the square lattice is defined by the following Hamiltonian:
\begin{equation}
     	H=-J_{x}\sum_{<i,j>_{x}}S_{i}S_{j}-J_{y}\sum_{<i,j>_{y}}S_{i}S_{j} +D\sum_{i}S_{i}^2 -h\sum_{i}S_{i}  
     \end{equation}
where $S_{i}= \pm 1,0, i=1,2 \cdots N$,  $N$ represents the total number of sites and $J_{x(y)}$ is  
the exchange interaction between sites along the two directions. $D$ is a single-spin anisotropy parameter and $h$ is an external magnetic field. Fig.~\ref{fig:figure3} (a) shows the ground-state phase diagram of the BC model. The temperature parameter $T/J_y$ is as low as 0.1.
The color characterizing different phases is obtained by the value of {$\sum S_i/N + |S_1- S_2|$.

In the next sections,  we use different classical phases to test our  SN-DCN method for the $h=0$ BC model and the extended SN-DCN method with two common nodes for the $h\ne 0$ BC model, respectively.

\subsection{The SN-DCN method}

\begin{figure}[t]
  \centering
  \includegraphics[height=3.9cm,width=6.cm]{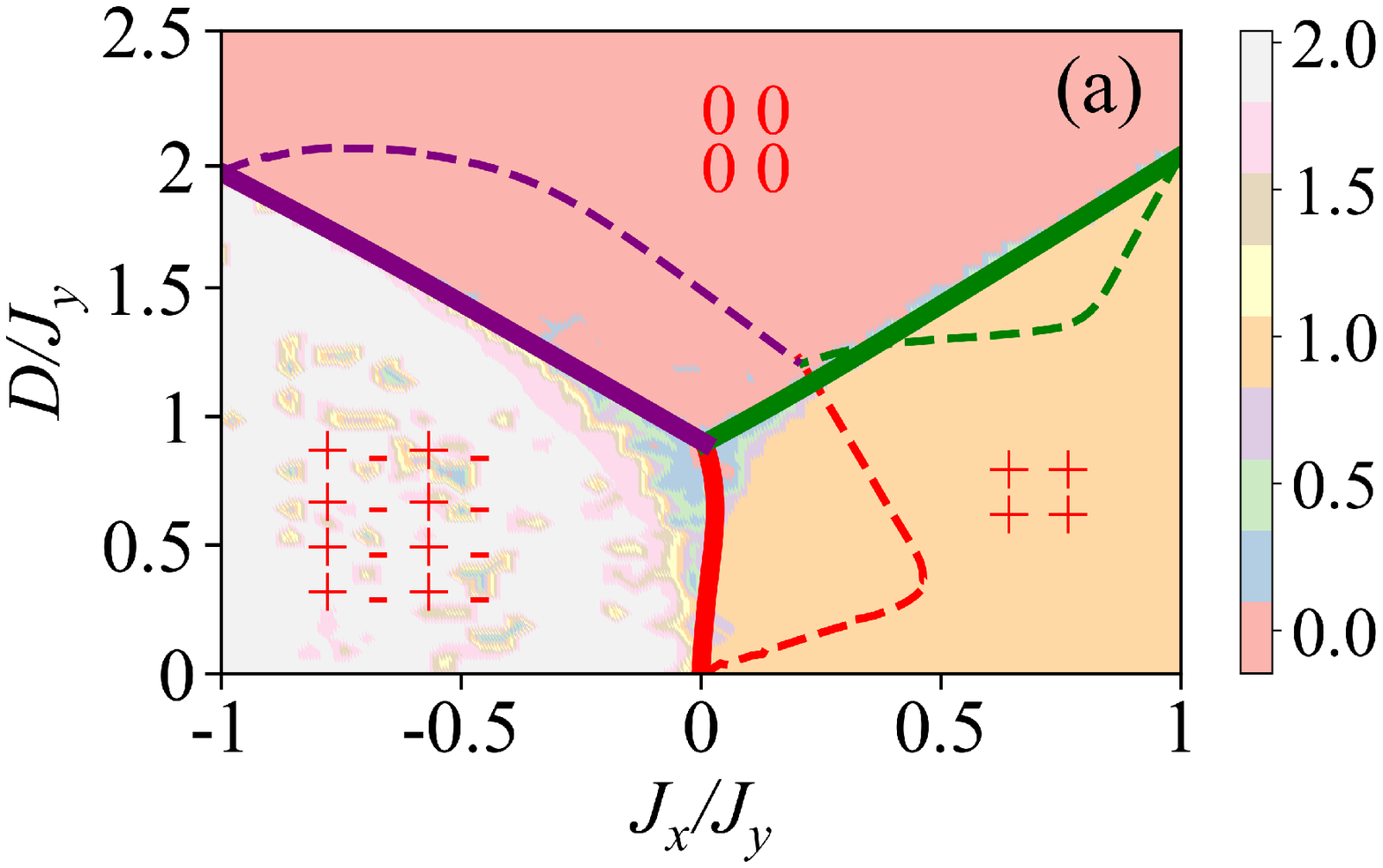}
  \vskip -0.12 cm
  \includegraphics[height=3.1cm,width=4.cm]{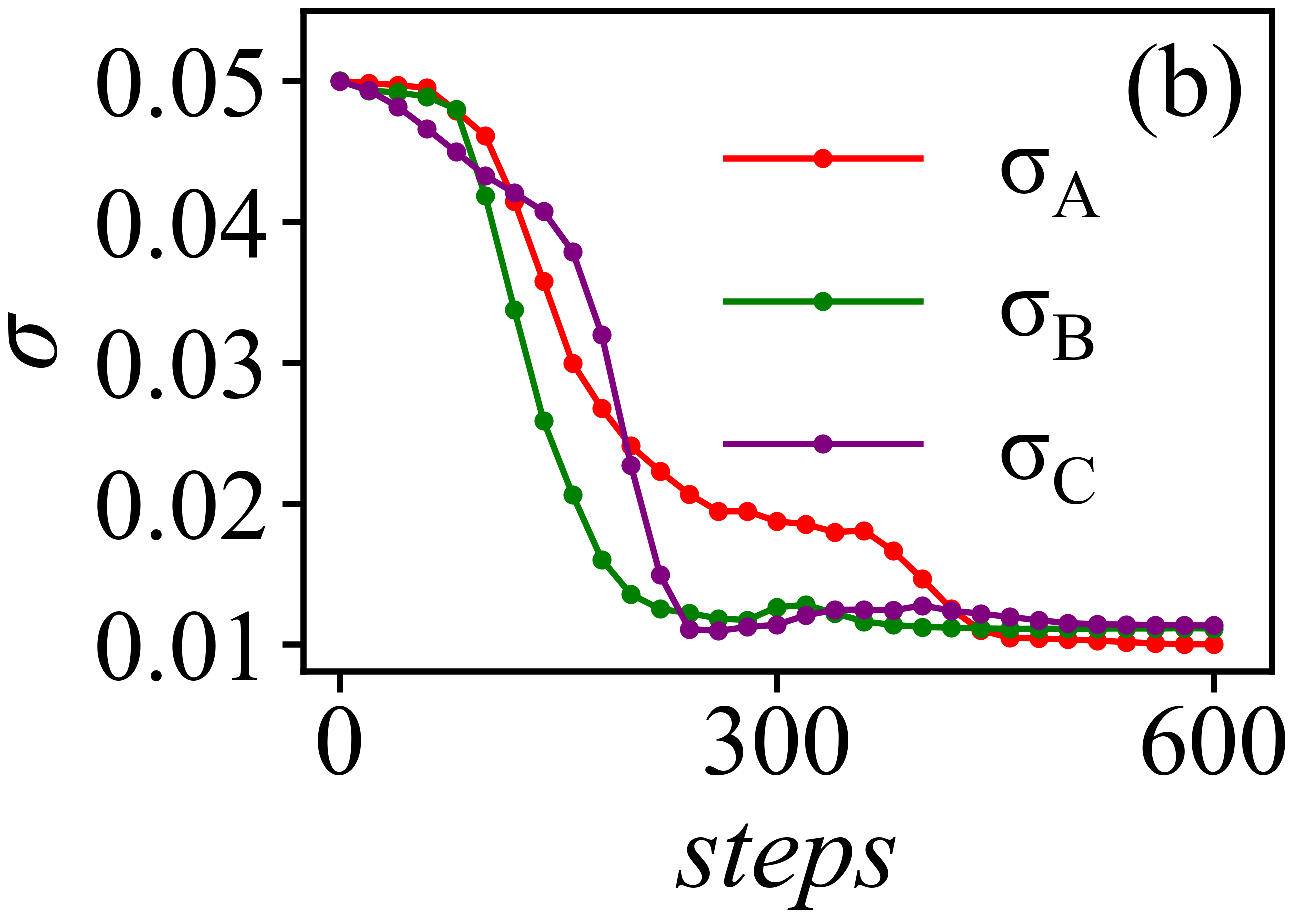}
  \hspace{-0.1 cm}
  \includegraphics[height=3.1cm,width=3.8cm]{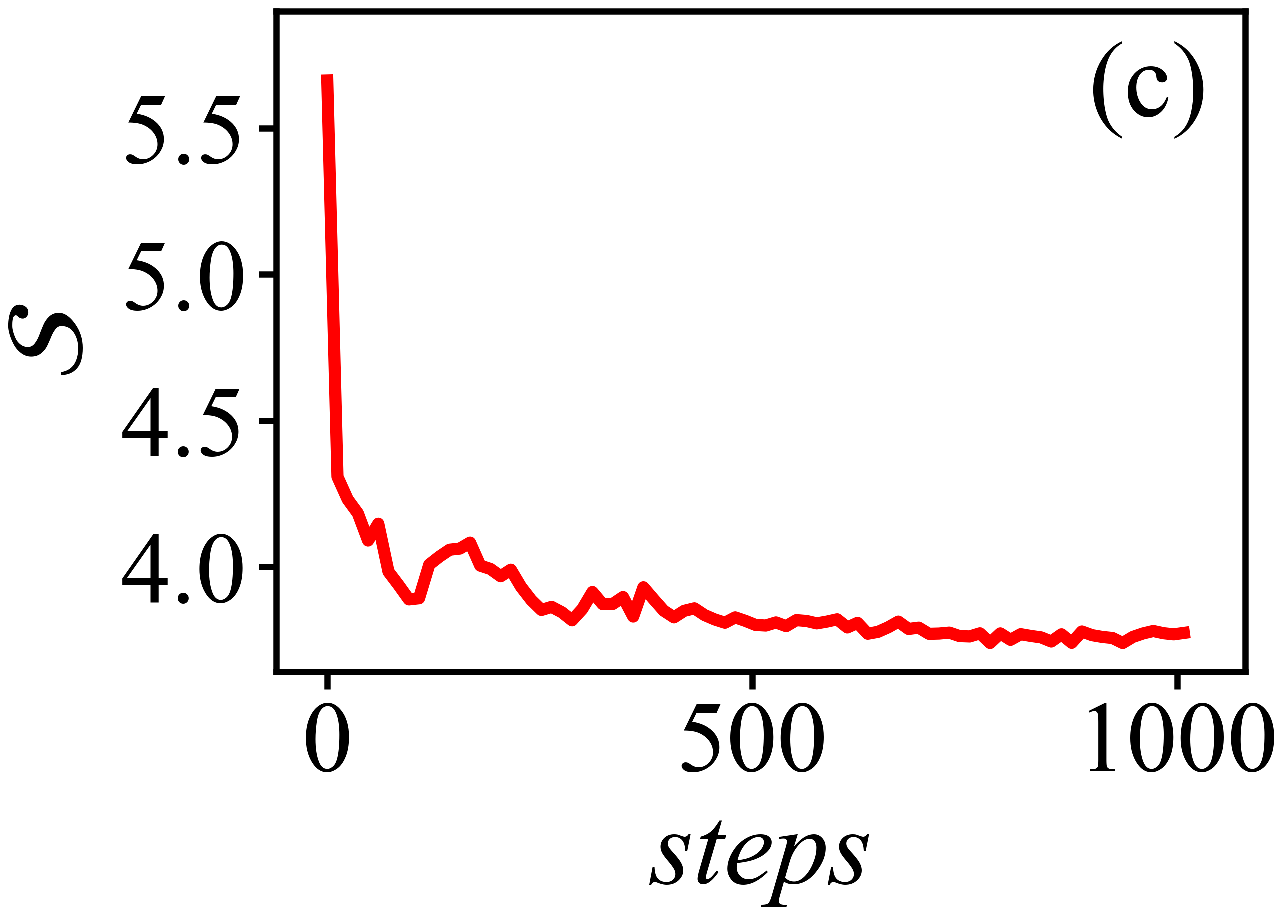}
  \vskip -0.08 cm
  \includegraphics[height=3.6cm,width=7.8cm]{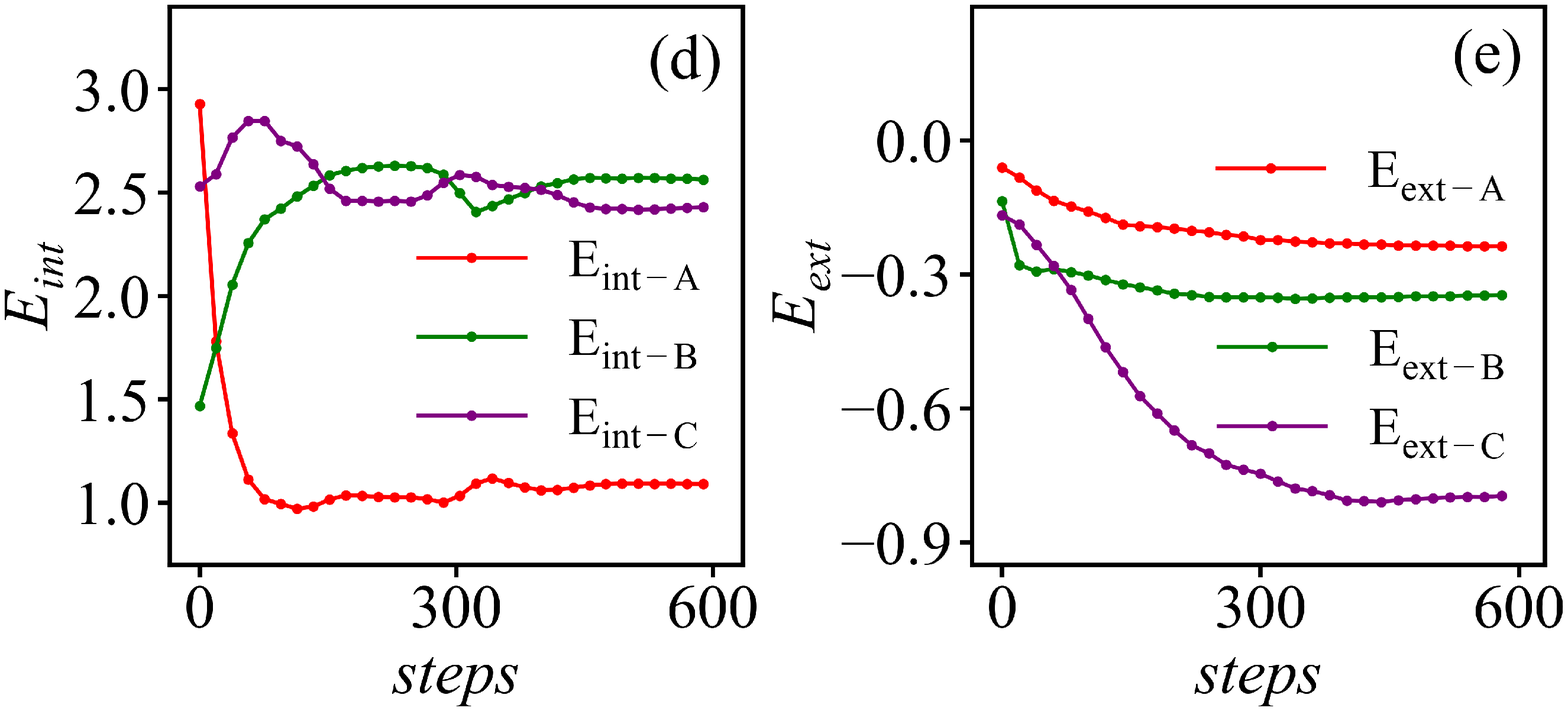}
  
  \caption{The results of applying  the SN-DCN method. (a) The phase diagram of the BC model with $h=0$, initial snakes ( dashed lines), and finial snakes (solid lines). (b) The average $\sigma$ versus  iteration steps. (c)  Total  $\mathcal{S}$  versus steps. (d)-(e) $E_{ext}$ and $E_{int}$  over steps.  These quantities converge very well. }
\label{fig:figure3}
\end{figure}

A simple SN  contains three snakes, which have a common node, as shown 
 in Fig.~\ref{fig:figure3} (a).
 %{\textcolor{blue}{Therefore, we call it  the one common node SN model.
 The initial  snakes are represented by dashed lines and the final snakes are marked by solid lines. Snakes $A$, $B$, and $C$ are marked in red, green, and purple, respectively.
The topology of the SN model is consistent with the boundaries between the three phases, which are the ferromagnetic, superantiferromagnetic and paramagnetic phases. The
configurations in a 4-site cell are  $\begin{smallmatrix} ++\\++ \end{smallmatrix}$, $\begin{smallmatrix} +-\\+- \end{smallmatrix}$ and $\begin{smallmatrix} 00\\00 \end{smallmatrix}$, respectively.
The parameters for the SN-DCN are listed in Appendix~\ref{sec:par1}.
 
According to Eq.~(\ref{eq:dyn}), during the updating process,  the width of the snake $\sigma$, the cross entropy cost $\mathcal{S}$, the external energy $E_{ext}$, and the internal energy $E_{int}$ of the snakes are also recorded separately to ensure that the snakes meets the mechanical balance.

In Fig.~\ref{fig:figure3} (b), the unit widths $\sigma_A$, $\sigma_B$, $\sigma_C$ of three snakes are shown.  The width for each node is updated independently, so the average width of all nodes for each snake is given here and these values converge from a value of 0.05 to about 0.01.
  %So we counted the average value of $\sigma$. Once the phase change point is within the snake's perception range, the $\sigma$ begins to plummet. This is to ensure that the snake locates the phase change point more quickly and accurately to ensure convergence. 
   
In Fig.~\ref{fig:figure3} (c),  the total cross entropy cost $\mathcal{S}$
 is convergent. 
{$\mathcal{S}$ is a function of the guessed boundary $\lambda_g$ and the unit width $\sigma$ of the snake. The DCN is used to find the minimum $\mathcal{S}$  and the corresponding $\lambda_g$ and $\sigma$ using the gradient descent method. 
Meanwhile, the values of $E_{ext}^{A,B,C}$ and $E_{int}^{A,B,C}$ also converge as shown in Fig.~\ref{fig:figure3} (d) and (e).
%According to Eq.~(\ref{eq:equation7}), the $E_{int}$ is related to the shape of the snake. 
The stability of the $E_{int}$ indicates that the shape of snakes no longer changes.  According to Eq.~(\ref{Eimg}), the value of pixel $I(C)$  
is replaced by  order parameters, i.e., the colors shown in Fig.~\ref{fig:figure3} (a). 
 The closer the snake is to the real phase boundary, the smaller the external energy is. 
These results show that the SN-DCN can be applied to a phase diagram with  three phases.

\subsection{The distinct initial topology of SN-DCN}
\label{sec:appendixB2}

In the previous subsection, 
the topology of the initial SN is consistent with  the true phase boundaries. Here we  discuss the correct results that are obtained with the wrong initial topology, i.e., the topology of the initial SN is different from the topology of the real boundaries.

\begin{figure}[htb] 
  \centering
  \includegraphics[height= 3.6cm,width=8.4cm]{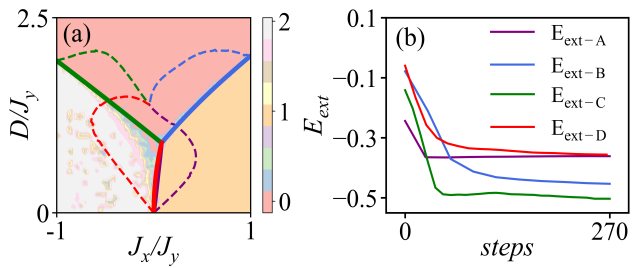} 
  \caption{(a) The SN-DCN containing four snakes is used to detect three boundaries. The initial SN are dashed lines and the final  are  solid lines. (b) The external energy of the two snakes overlaps finally. }
\label{fig:figure12}
\end{figure}

In Fig.~\ref{fig:figure12} (a),  initially, there are four snakes (dashed lines), but there are only three boundaries.  Snakes $A$, $B$, $C$ and $D$ are marked in purple, blue, green and, red, respectively.
After updating, eventually, the two snakes (red  and purple) overlap to a single true phase boundary. This shows that even with an extra snake, our SN-DCN can still find the true phase boundary correctly.
To further check whether or not other properties overlap, when the two snakes positions converge, the energy of $snake_A$ and $snake_D$ overlap, as shown in Fig.~\ref{fig:figure12} (b).
\subsection{The extended SN-DCN method}
\begin{figure}[hbt]
 \includegraphics[height=4.1cm,width=6.5cm]{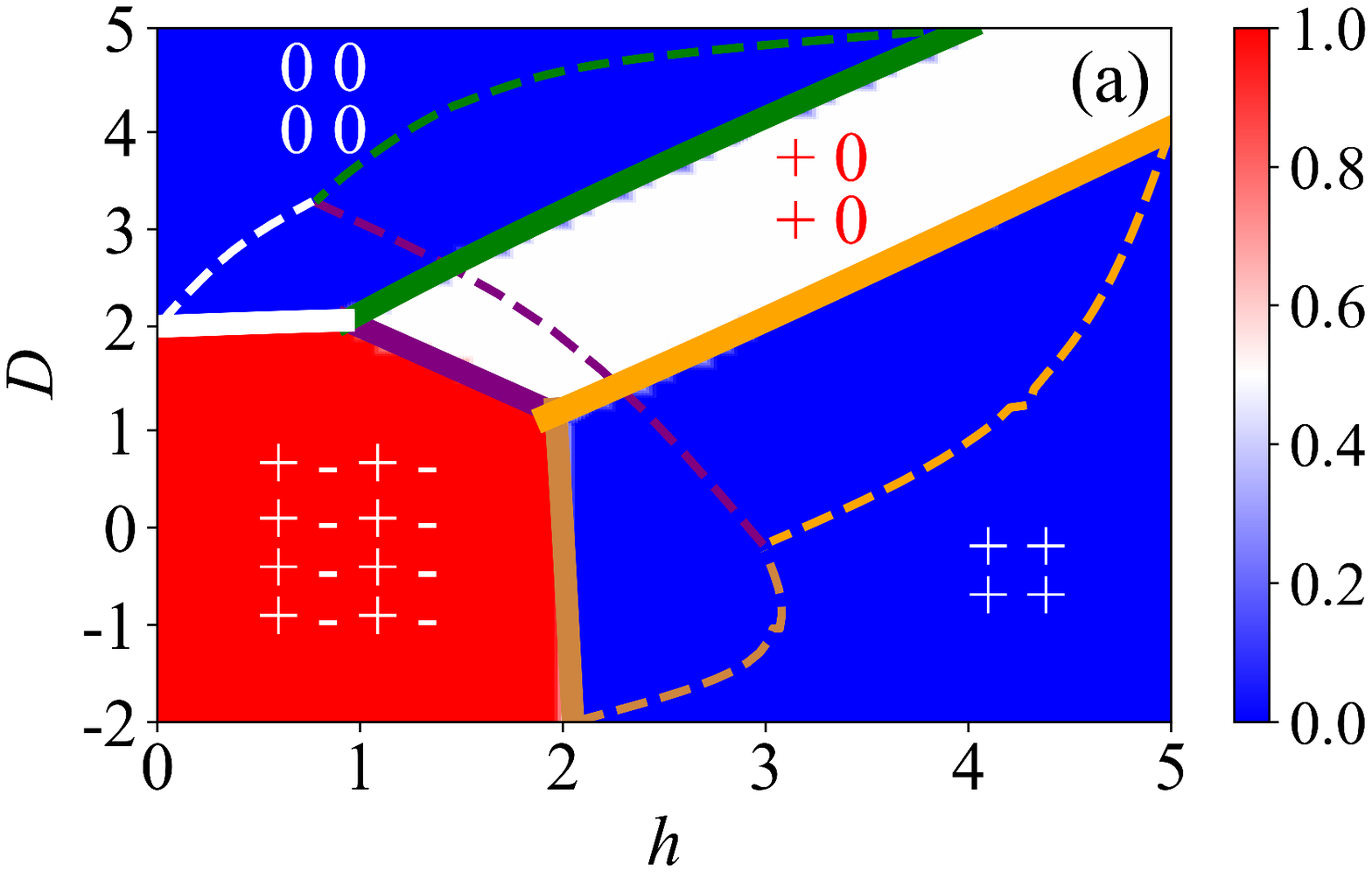}
 \vskip -0.16 cm
 \includegraphics[height=3.95cm,width=4.3cm]{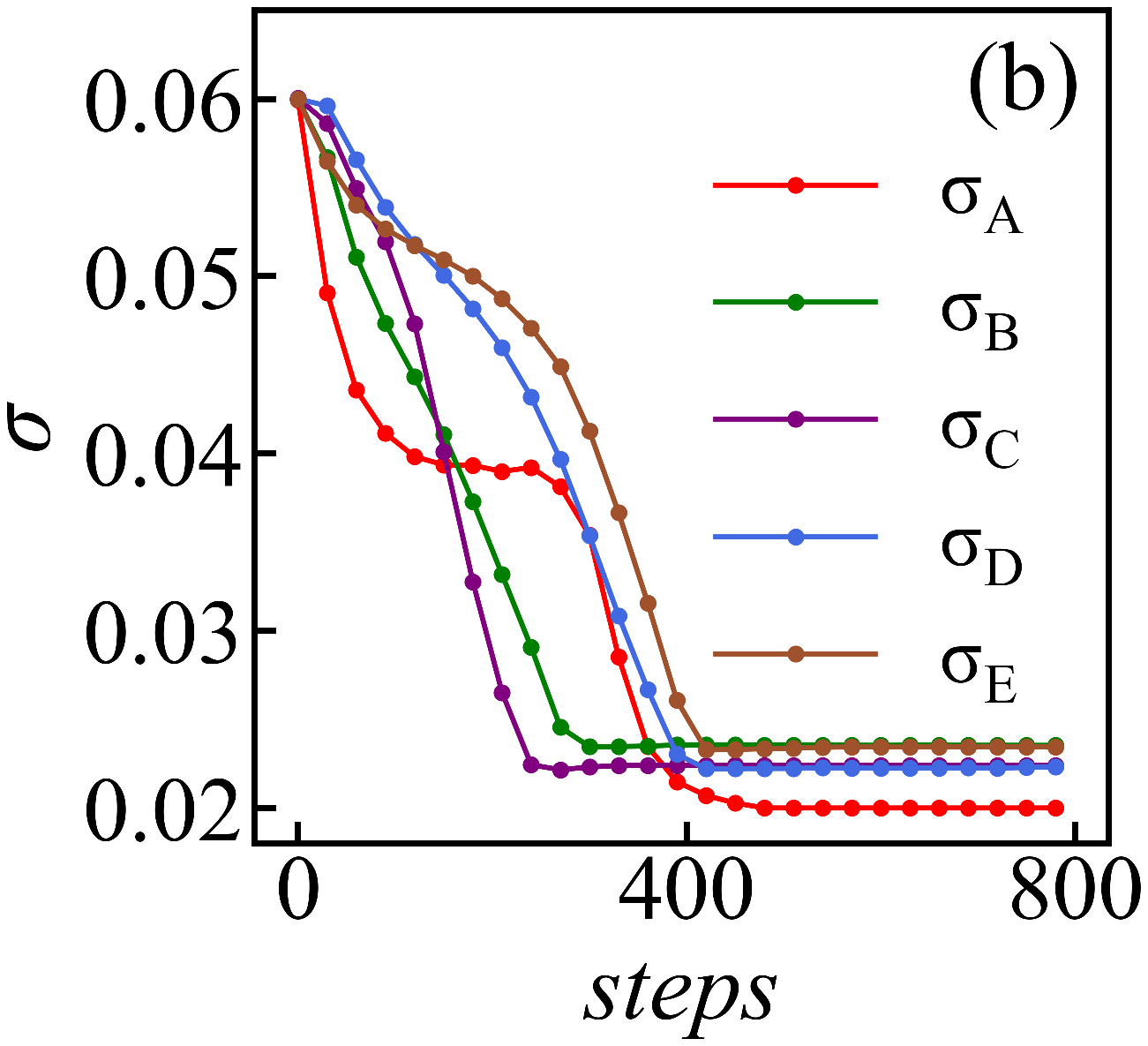}
 \hspace{-0.14 cm}
 \includegraphics[height=3.9cm,width=4.05cm]{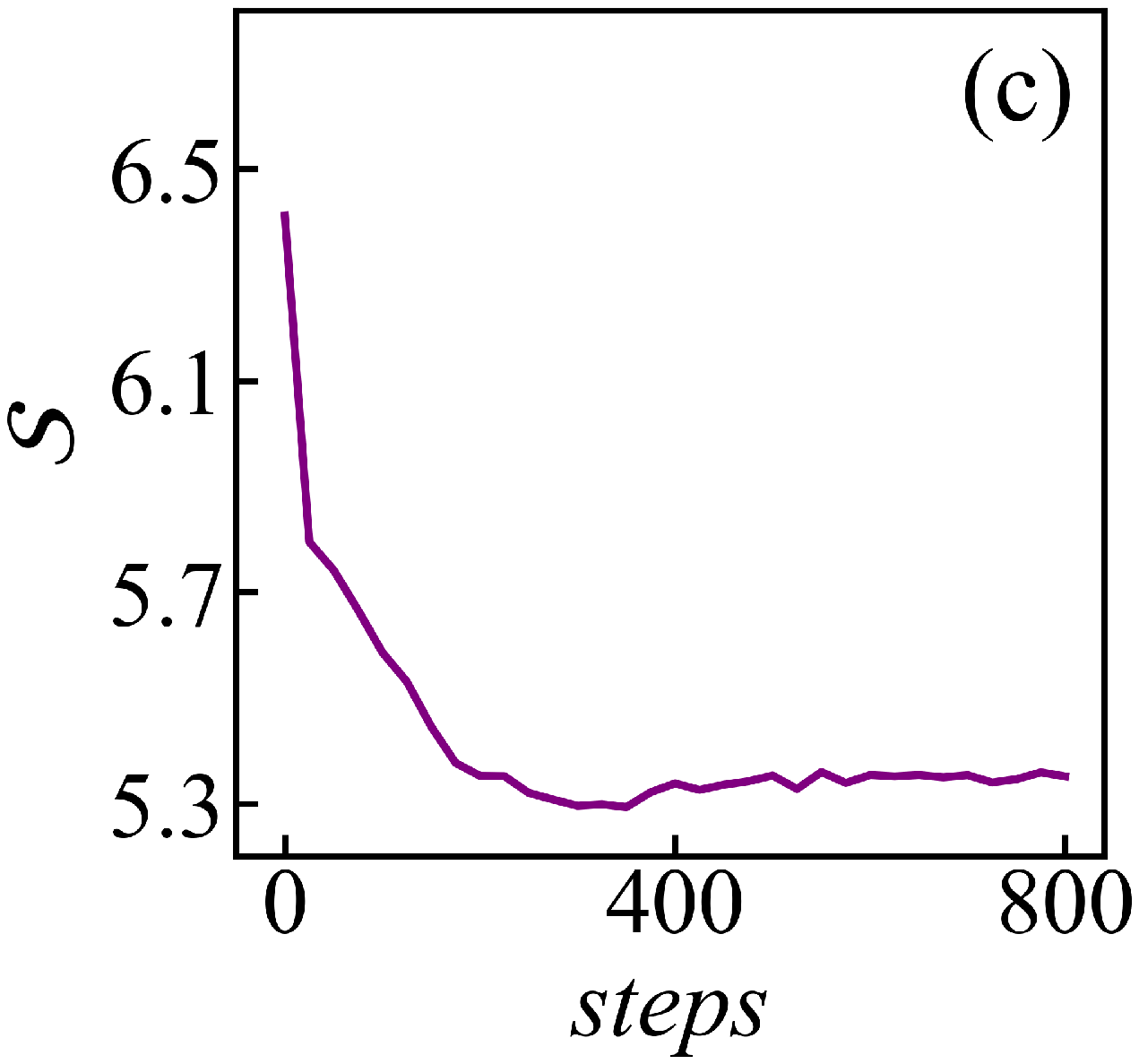}
 \vskip -0.1cm
 \includegraphics[height=3.9cm,width=4.1cm]{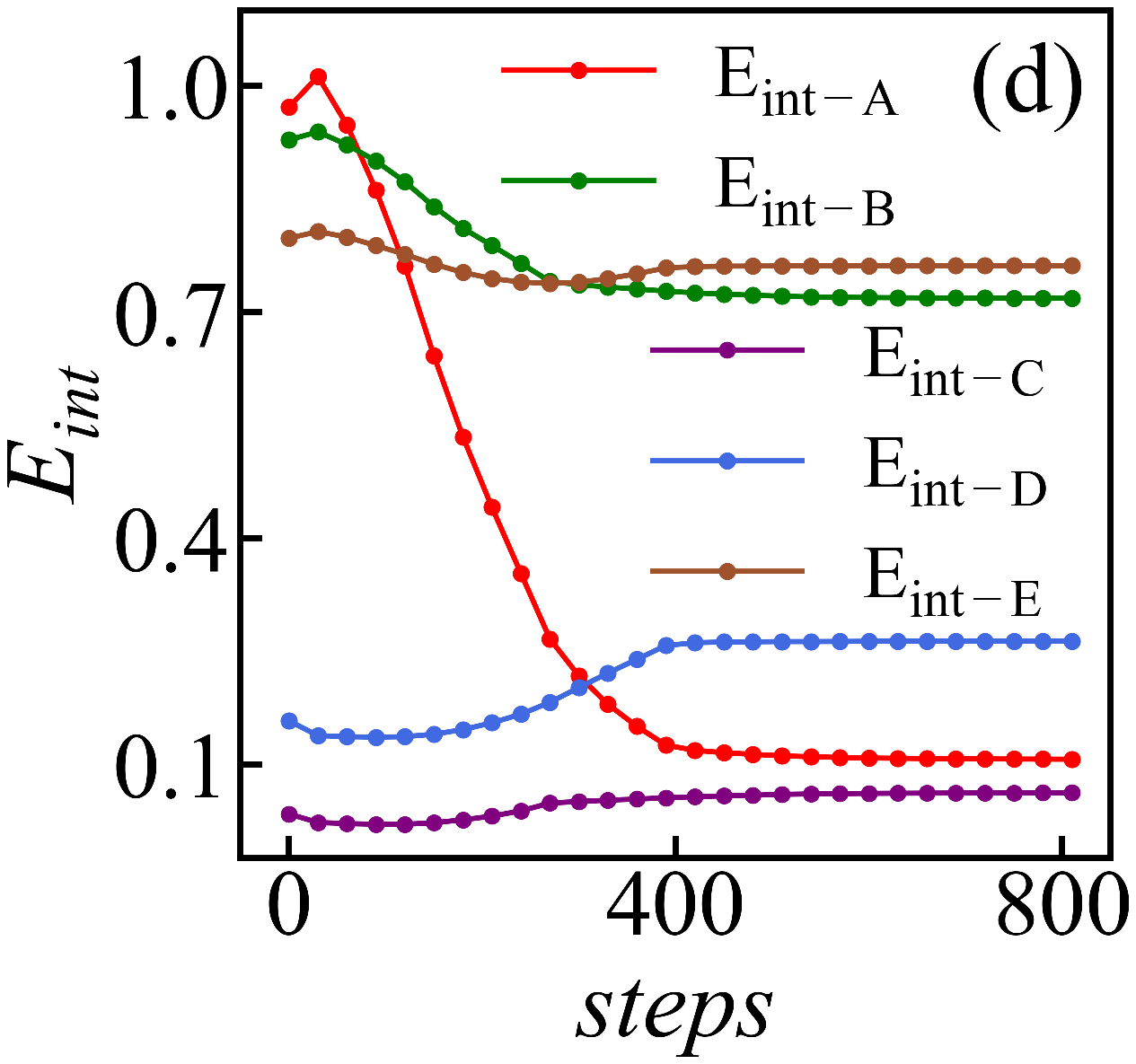}
 \hspace{-0.25 cm}
 \includegraphics[height=3.9cm,width=4.2cm]{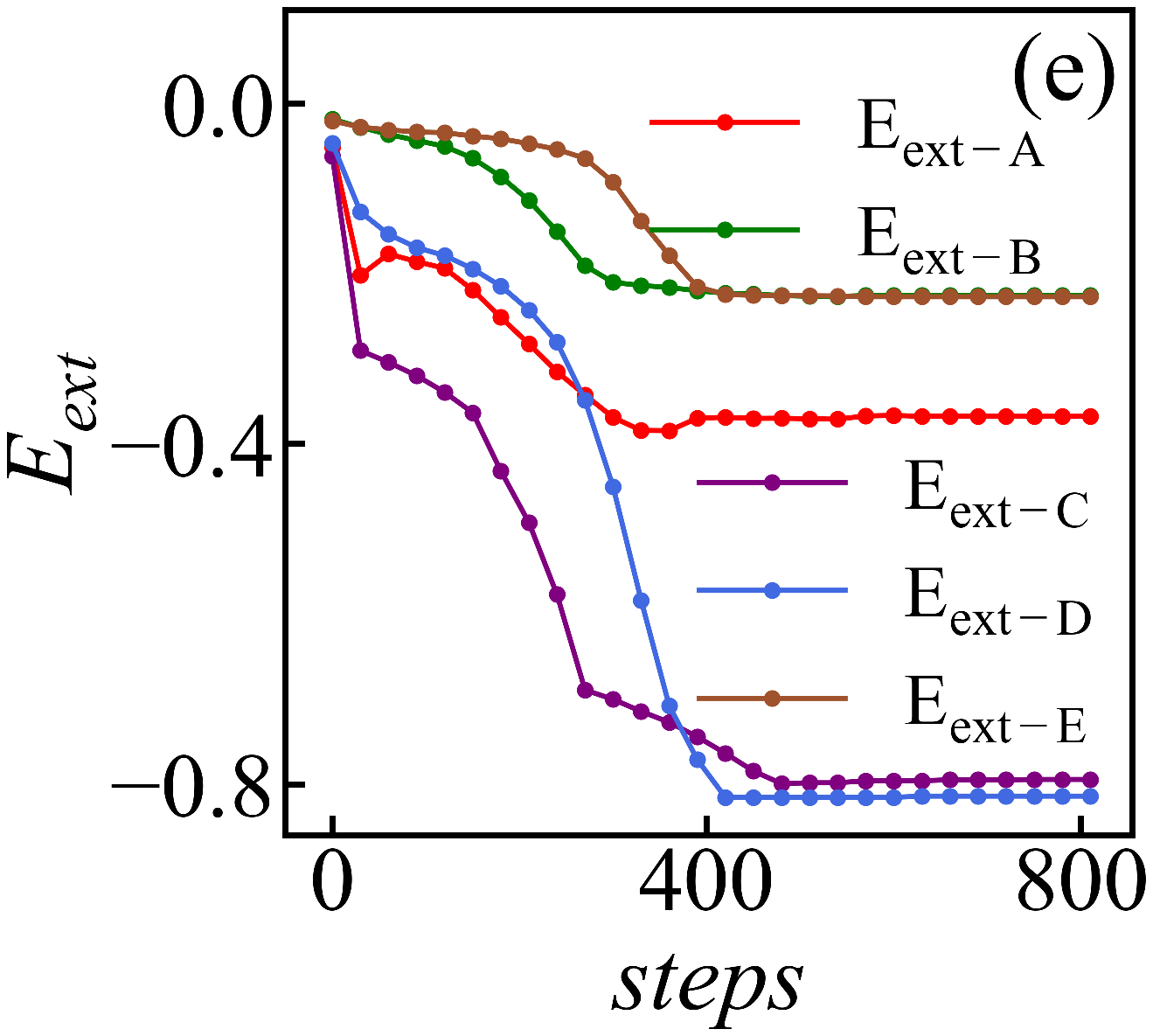}
 \caption{The results of applying  the extended SN-DCN method. (a) The ground state phase diagrams of the BC model in the plane $D$ for $h$ for $J_{x}/J_{y}=-1$, initial snakes ( dashed lines) and finial snakes (solid lines). (b) The average $\sigma$ versus  iteration steps. (c)  Total  $\mathcal{S}$  versus steps. (d)-(e) $E_{ext}$ and $E_{int}$  over steps.  These quantities converge very well. }
\label{fig:figure4}
\end{figure} 

To test the generality of the SN-DCN method,   the two movable common nodes are introduced. It can detect boundaries between four phases. 
Fig.~\ref{fig:figure4} (a) shows the ground state phase diagram of the BC model with $J_{x}/J_{y} = -1$.  It contains four phases, whose configurations are $\begin{smallmatrix} ++\\++ \end{smallmatrix}$,  $\begin{smallmatrix} +-\\+- \end{smallmatrix}$, $\begin{smallmatrix} +0\\+0 \end{smallmatrix}$, and $\begin{smallmatrix} 00\\00 \end{smallmatrix}$. 
The color is obtained by $\sum_{n=1}^{4} |S_i- S_{in}|/4$, where $S_{in}$ are the
spins located at the neighboring lattice sites.

The initial snakes are illustrated by the dashed lines and  then the phase boundaries are detected by the final snakes marked by solid lines in different colors. During the updating process, the quantities $\sigma$,  $\mathcal{S}$,  $E_{int}$ and  $E_{ext}$ are shown in Fig.~\ref{fig:figure4} (b)-(e), and all of them can be convergent, which means that the SN model is {\it extendable}.
The parameters for obtaining Fig.~\ref{fig:figure4}} list in
the Appendix~\ref{sec:par2}.

\section{The balloon force and its application}
\label{sec:balloon}

\subsection{The motivation of introducing the balloon force }

\begin{figure}[htb]
 \includegraphics[height=4.1cm,width=4.1cm]{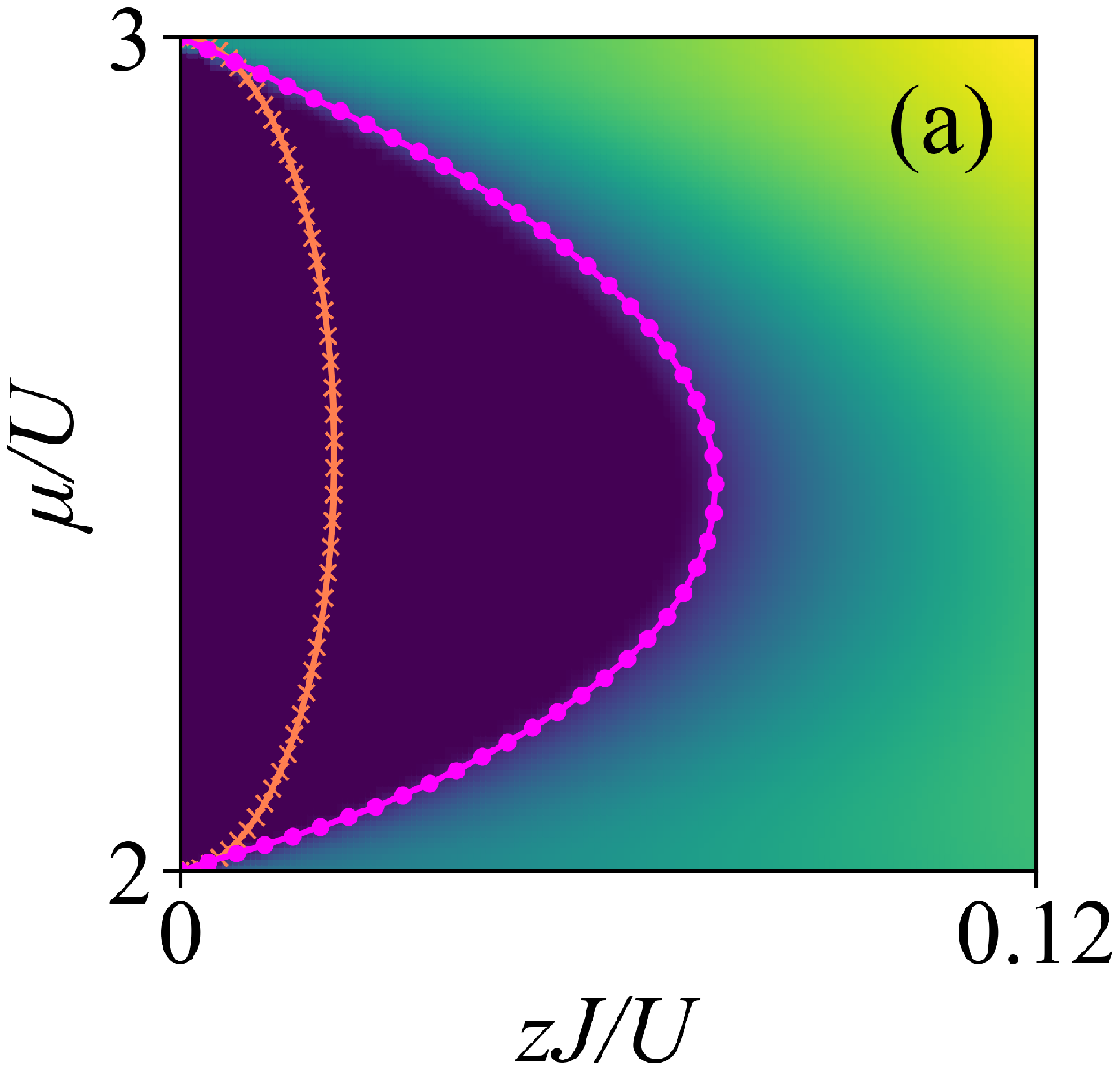}
 \hspace{-0.25 cm}
 \includegraphics[ height=4.1cm,width=4.1cm]{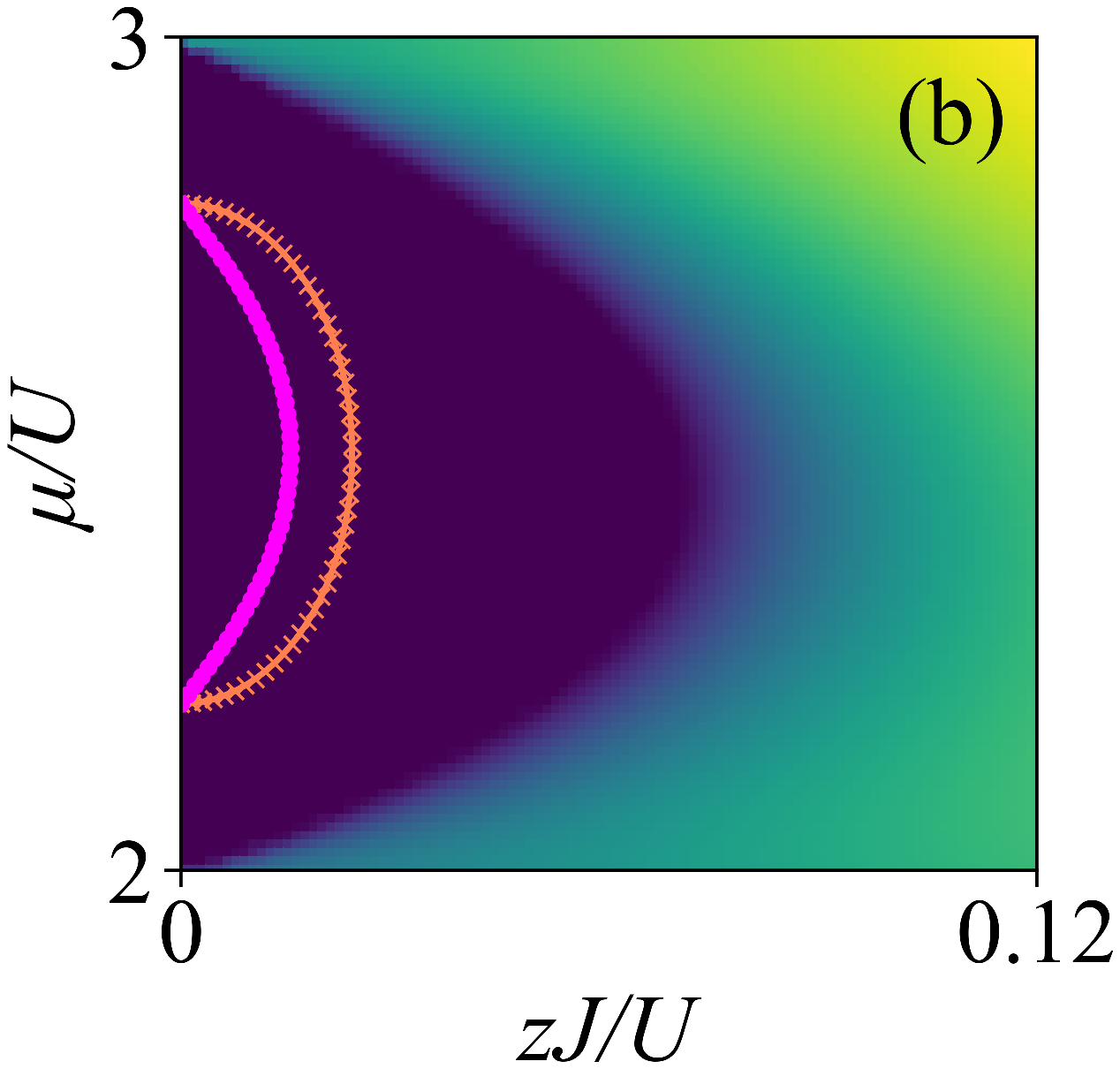}
 \caption{ (a) Without the balloon force, 
the DCN can help find the correct phase boundary (purple) if the initial snake (red) is set at the correct initial position. 
 (b) The DCN fails to  find the correct phase boundary (purple) if the initial snake (red) is set at the wrong initial position. } 
 \label{fig:figure5}
\end{figure}

The balloon force is  inspired by the field of computer image processing~\cite{COHEN1991211}. It is used to solve the problem that when the initial snake is far from the target contour, the snake cannot feel the image force and cannot move.
For physical systems, the pixel points in the image are replaced by thousands of physical configurations.
It is not clear whether balloon force can help detect the boundary of the phase. 

We use the phase diagram of the BH model to illustrate the effect of the balloon force. The Hamiltonian of the BH model is expressed as~\cite{BH}:
\begin{equation}
 	H=-J\sum_{<i,j>}(b_{i}^{\dag}b_{j}+b_{j}^{\dag}b_{i})+\sum_{i}(\frac{Un_{i} (n_{i}-1)}{2}-\mu n_{i}),
\end{equation}
where $\mu$ is the chemical potential, $J$ and $U$ are the boson hopping energy and on-site interaction, respectively.  $b$ and $b^{\dag}$ are the boson creation and annihilation operators, respectively. $n_i$  represents the particle number operator of the site $i$.
Using the mean-field  approximation~\cite{MF}, the order parameter $\psi = < b_{i}^{\dag}>= <b_{i}>$ can be introduced to describe the superfluid  and insulated phases.

In Ref.~\cite{liuprl}, the initial snake first encloses the target contour, and then it gradually shrinks to the target contour.
Here, as shown in Fig. ~\ref{fig:figure5} (a), the snake is initialized at a different location marked by the red symbols, i.e., within the target contour.
The ending points of the initial snake are fixed at the ends of the axis $zJ/U = 0$. The purple line represents the snake after convergence.

However, in Figs.~\ref{fig:figure5} (b) the initial snake is  fully immersed in the insulating phase marked in blue, and the snake is finally located near the initial position. Only the shape made a small change under the action of internal forces.
The reason is that  the snake hardly feels the external force which is provided by cross entropy cost $\mathcal {S}$. 
According to Eq.~(\ref{eq:equation2}), the sampled data $d(\lambda)$ from the parameter marked with green symbols in Fig.~\ref{fig:FIG1} (c), is from the same phase, $\mathcal {S}$ is a constant because the outputs 
$p_{A,B}^\mathcal{G}$, and $p_{A,B}^\mathcal{N}$ of $\mathcal{G}$ and $\mathcal{N}$ do not change. 
According to Eq.~(\ref{eq:dyn}b),  the nodes of the snakes  cannot move.

To solve this limitation that the initial snake position must be close to the real boundary, we introduce the BF-DCN method.  Here, a {\it decaying} balloon force is defined as:
\begin{equation}
	{\bf F}_{balloon}=\kappa{\bf n}(C) \label{eqbf} \end{equation} where $\kappa$ decays with the iteration steps $\kappa=\kappa_0 a^{-steps}$, and  ${\bf n}(C)$ means the normal direction of the snakes. The direction ${\bf n}(C)$ of the ${\bf F}_{balloon}$ outward along the normal direction is positive and inward is negative.

To visualize the iterative process more clearly, 
 the average distance $D$ is defined as:
\begin{equation}
     	D=\frac{\sum_{i=1}^N \sqrt{|x_{i}^{t}-x_{i}^{\infty}|^2+|y_{i}^{t}-y_{i}^{\infty}|^2\quad}}{N},
     \end{equation}
 where $(x_i^t, y_i^t)$ and  $(x_i^{\infty}$, $ y_i^{\infty})$ are the coordinates at time $t$ and the final coordinates, respectively. $N$ represents the number of nodes in each snake. For convenience, 
 We follow the custom of Ref.~\cite{liuprl} and normalize the range of coordinates in the physical parameter space.

\subsection{Force analysis of snake nodes}

\begin{figure}[tbh]
 \includegraphics[height=4.1cm,width=4.1cm]{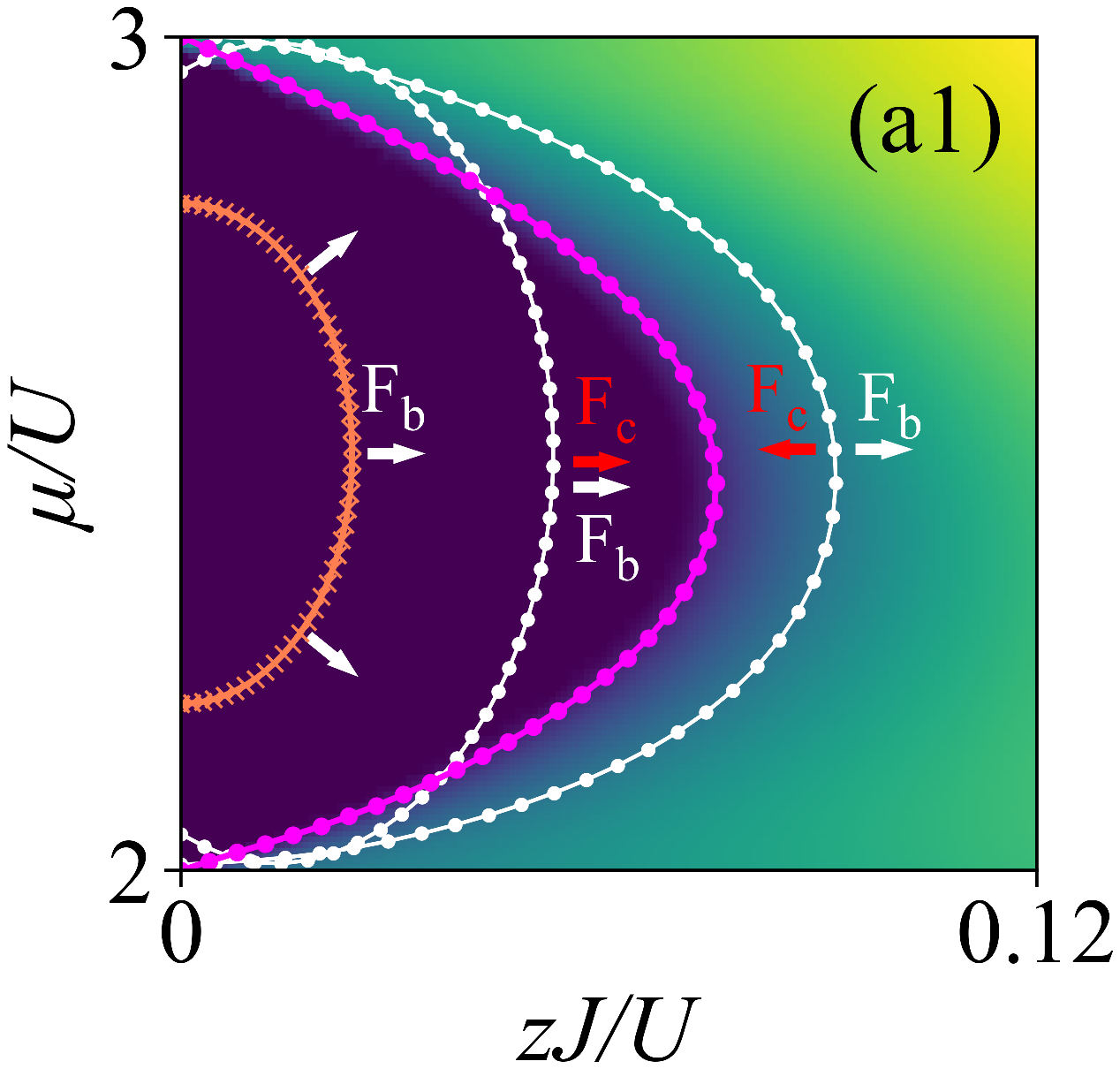}
 \hspace{-0.25 cm} 
 \includegraphics[height=4.0cm,width=4.05cm]{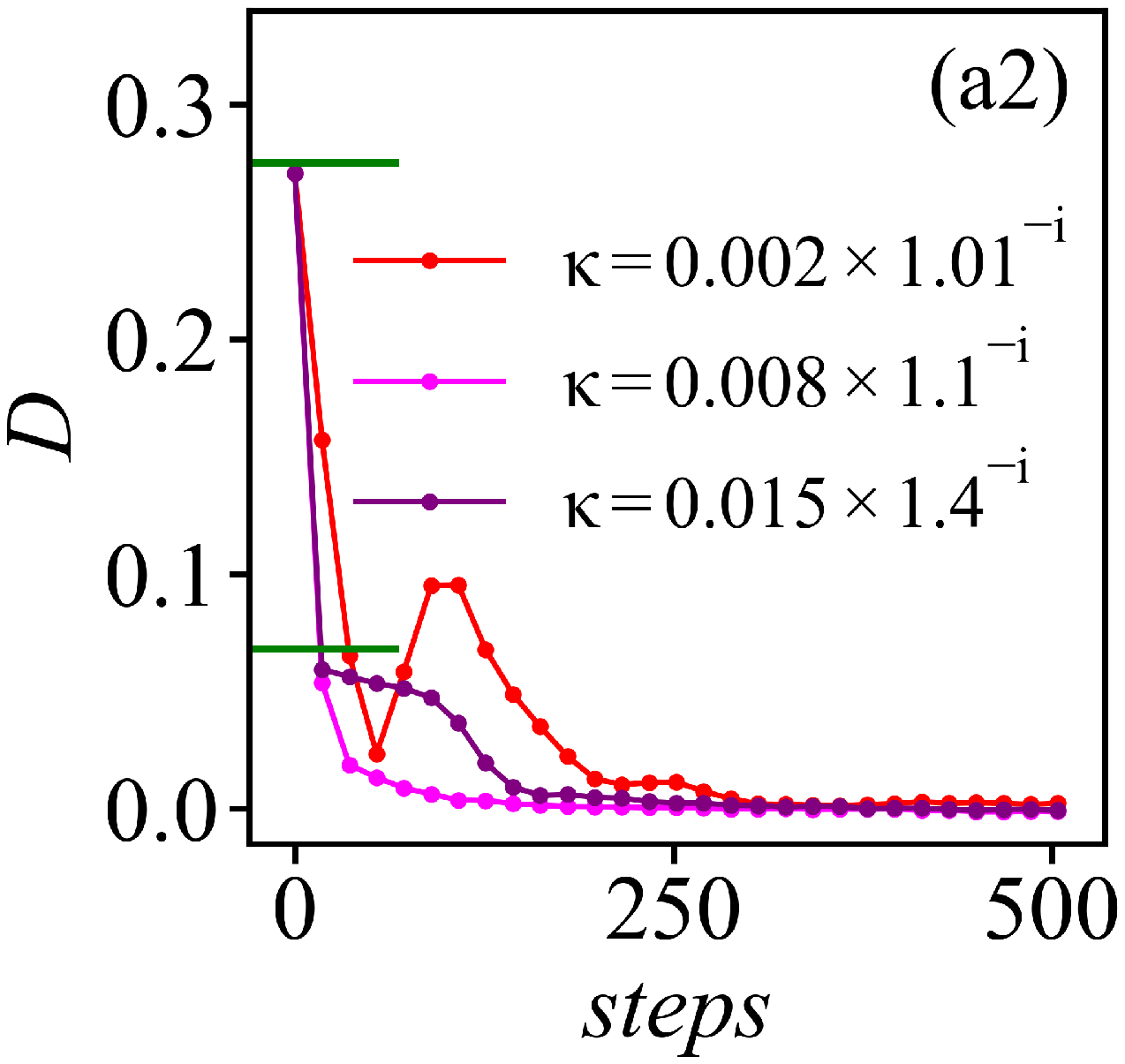}
 \vskip -0.15 cm
 \includegraphics[height=4.1cm,width=4.1cm]{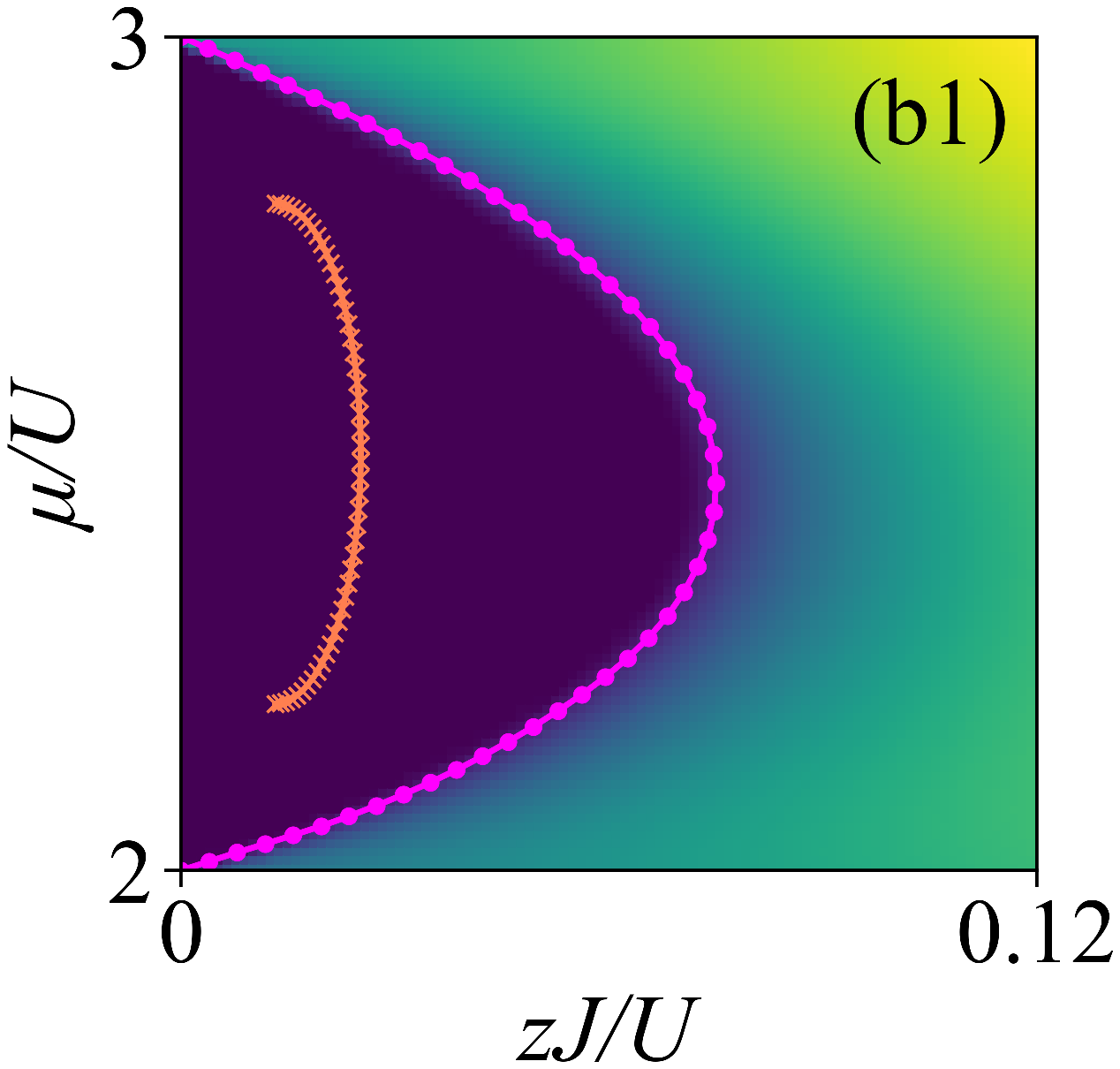}
 \hspace{-0.25 cm}
 \includegraphics[height=4.0cm,width=4.05cm]{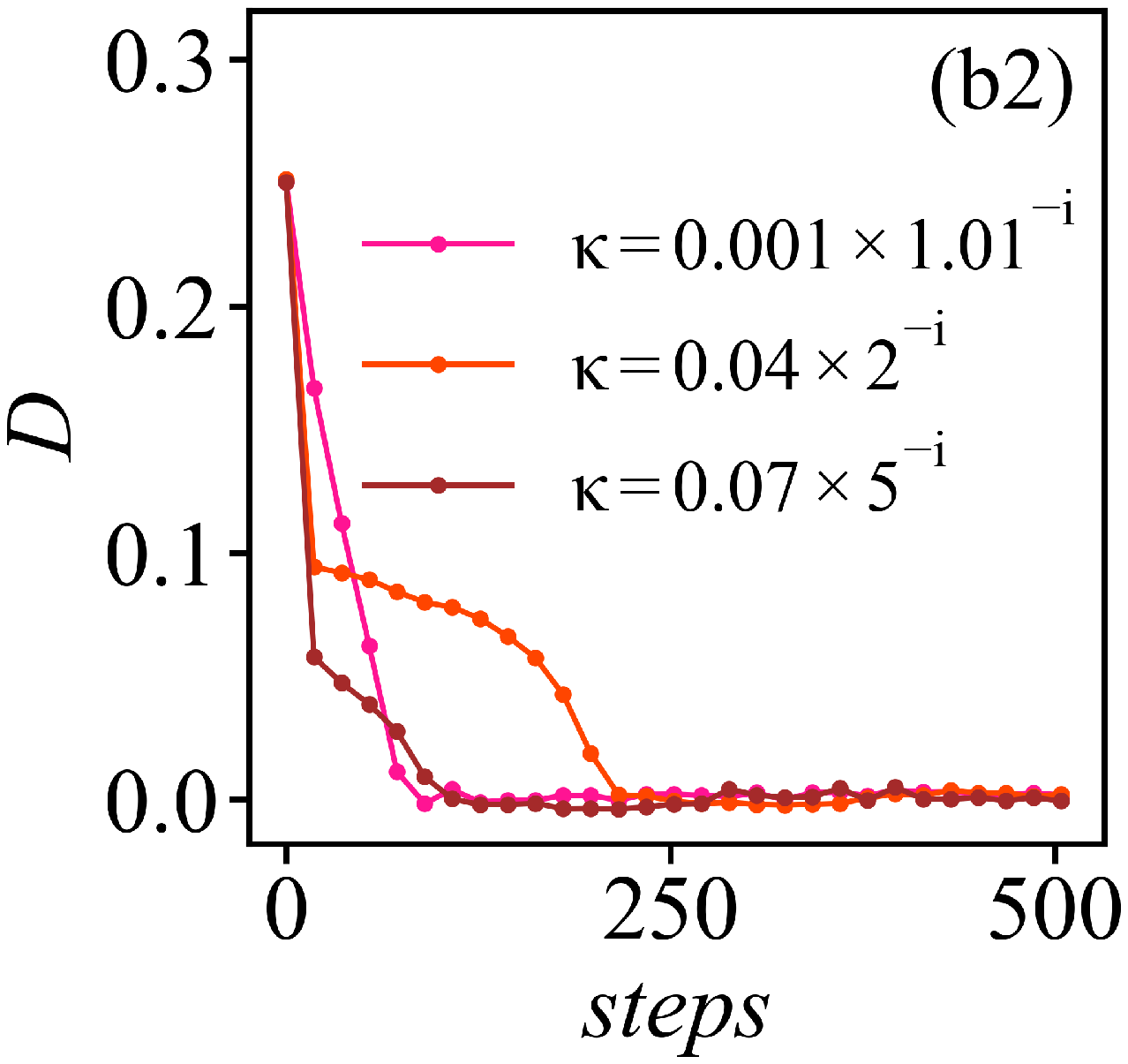}
 \caption{Force analysis of snakes 
  with balloon force, (a1)  the initial snake (red) with only balloon force $F_{b}$,  and the final snake (purple). Two possible positions of snakes (white lines) with forces, $F_{cost}$ and $F_{balloon}$, simplified as $F_{c}$ and $F_{b}$. (a2)  $D$ versus steps for (a1). (b1) The initial snake is completely immersed in the insulated phase. (b2) $D$ versus steps for (b1).}
\label{fig:figure6}
\end{figure}

To understand how the balloon force works, a force analysis of the snake is performed here. The total force $F_{tot}$ is composed of the internal force $F_{int}$ and external force $F_{ext}$, i.e.,  
\begin{equation}
    {\bf F}_{tot}={\bf F}_{int}+{\bf F}_{ext}, ~~~{\bf F}_{ext}={\bf F}_{cost}+{\bf F}_{balloon},
\label{eq:Equation19}
\end{equation}
where ${\bf F}_{cost}$  is the force introduced by the gradient descent method to find the minimum cross entropy cost, and its direction is the direction normal to the snake.
${\bf F}_{int}$ depends on parameters such as $\alpha$, $\beta$, etc., and only changes the appearance characteristics such as whether the snake is smooth or not, but not the overall position. Therefore the magnitude of ${\bf F}_{int}$ is not analyzed here. 

In Fig.~\ref{fig:figure6} (a1), the forces on four snakes are shown. 
The balloon forces are marked by 
white arrows and labeled as ${\bf F}_b$, i.e., ${\bf F}_{balloon}$.  When using the balloon force, the position of the real phase boundary relative to the initial snake needs to be known. The sign of ${\bf F}_b$ cannot be varied in our approach. Under the action of ${\bf F}_{balloon}$, the initial snake marked in red begins to expand gradually to the right.

The snake located at other possible locations are also shown, where the snake marked with the white line on the left side perceives ${\bf F}_{balloon}$ and ${\bf F}_{cost}$ in the same direction, while the snake marked with the white line on the right side perceives ${\bf F}_{balloon}$ and ${\bf F}_{cost}$  in the opposite direction. The combined effect of the two forces confines the snake to the real phase boundary.  In Fig.~\ref{fig:figure6} (a2), between the two green lines,  the snake is moving fast. This is because, in the early stages, the balloon force has not decayed as much. In the final stages, $D\approx 0 $ means the snakes converge to the true phase boundary. 
 In Fig. ~\ref{fig:figure6} (b1), with the help of ${\bf F}_{balloon}$, at an  initial position with ${\bf F}_{cost}=0$, the snake is still able to iterate to the target position. The quantity $D$ is shown in Fig. ~\ref{fig:figure6} (b2).

 Here we provide a short argument why the snake is guaranteed to converge in with a  decaying force. As shown in Eq.~(\ref{eq:Equation19}), the external force  ${\bf F}_{ext}$ includes ${\bf F}_{cost}$ and ${\bf F}_{balloon}$, The function of ${\bf F}_{cost}$ is similar to the restoring force of a spring, dedicated to pulling the nodes of the snake back to the equilibrium position, or the true phase boundary, where ${\bf F}_{cost}=0$. 
 The balloon force should also be close to  0 at the phase boundary, otherwise,  if the balloon force is non-zero constant the snake can go beyond the true phase boundary.

\subsection{The super-parameters and the improvement  by the balloon force}

\begin{figure}[hbt]
 \includegraphics[height=3.6cm,width=7.9cm]{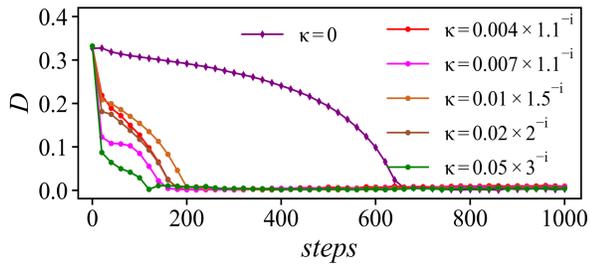} 
 \caption{ $D$ versus steps with ($\kappa\ne 0$) and without ($\kappa=0$) balloon force. Clearly, the  balloon force helps accelerate the convergence.}
 \label{fig:figure8}
\end{figure}

\begin{figure}[hbt]
 \includegraphics[height=3.1cm,width=7.9cm]{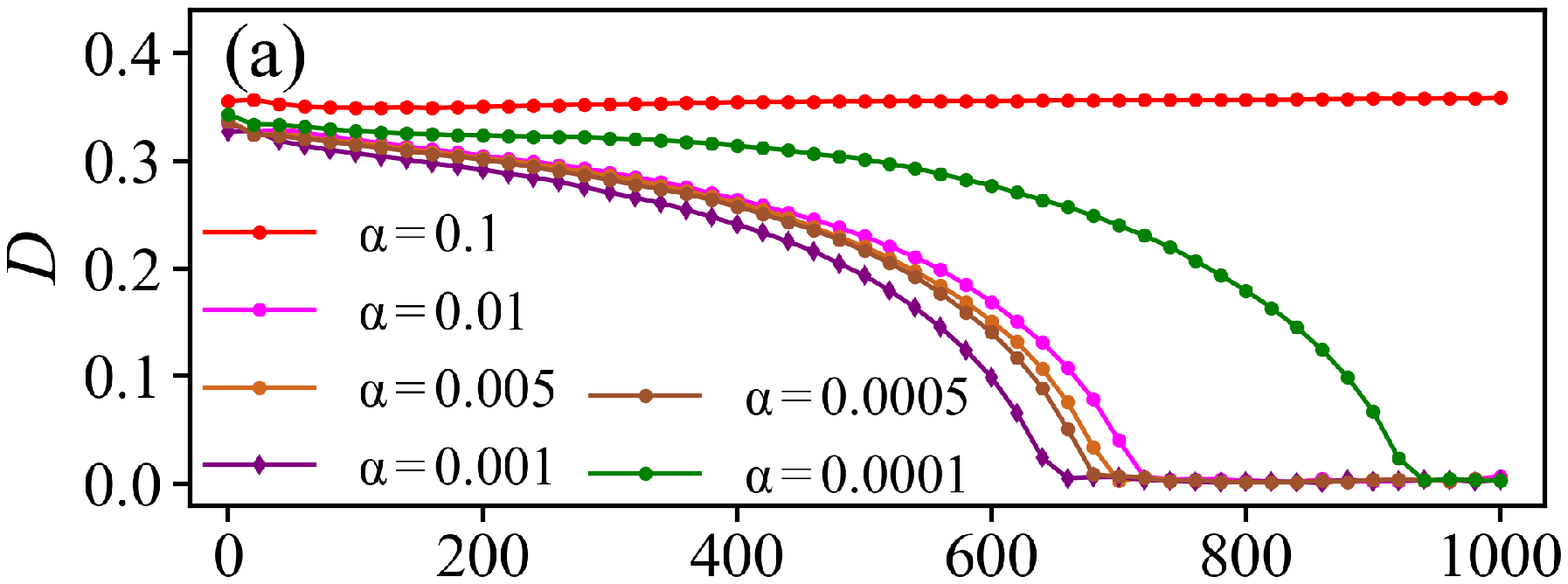}
 \vskip -0.16 cm
 \includegraphics[ height=3.1cm,width=7.9cm]{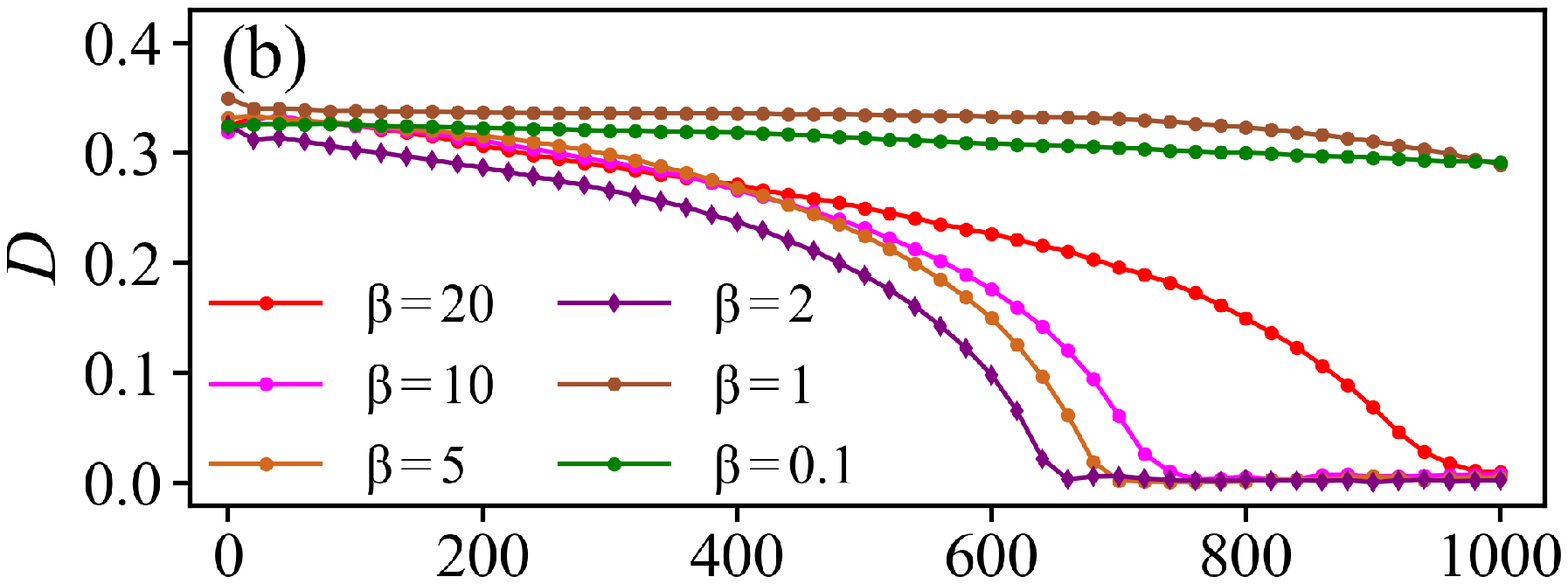}
 \vskip -0.16 cm
 \includegraphics[height=3.5cm,width=7.9cm]{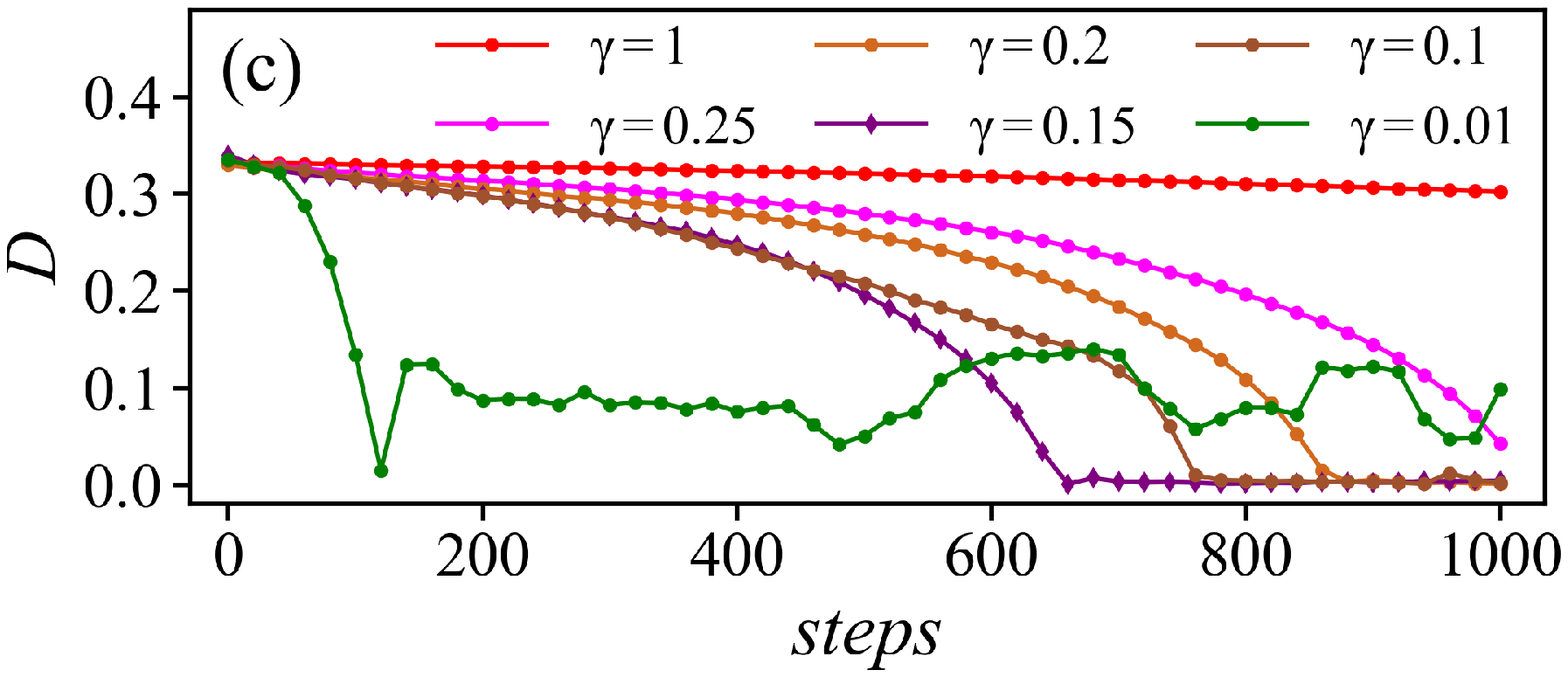} 
 \caption{ Without balloon force, $D$ versus steps for various parameters $\alpha$, $\beta$ and  $\gamma$. (a) Adjusting only $\alpha$,   (b)  adjusting only $\beta$,  (c)  adjusting only $\gamma$ .}
 \label{fig:figure7}
\end{figure} 

In Fig.~\ref{fig:figure8}, the data of  $D$ .vs. $steps$ show that   ${\bf F}_{balloon}$  reduce number of training step. For comparison purposes, the initial positions corresponding to the different data lines are the same.
$\kappa=0$, i.e.,  ${\bf F}_{balloon}=0$  results in a slow convergence effect with convergence steps to over 600. Other parameters are set to $\alpha = 0.001$, $\beta = 2$, and $\gamma = 0.15$.   These parameters correspond to the  fastest  convergence with ${\bf F}_{balloon}=0$. So we choose this set of data as a comparison for  ${\bf F}_{balloon} \ne 0$.

We also show the effects of  other parameters. 
We adjust many values of $\alpha$, $\beta$, and $\gamma$, and none of them are found to accelerate the convergence of $D$ more easily than the balloon force.
Without the balloon force, 
the  distances $D$ are shown versus iteration steps with different values of $\alpha$, $\beta$, and $\gamma$. 
In Fig.~\ref{fig:figure7} (a), $\alpha$ varies from 0.1 to 0.0001. The fastest parameter is an intermediate value of 0.001. 
The reason is that large $\alpha$  makes the snake straight and hinders bending. Small $\alpha$ leads to the curve being too easy to bend without being rigid.
It has notorious difficulty in determining the weights $\alpha$, $\beta$, and $\gamma$ associated with the smoothness constraint, reported in a review reference~\cite{rsnake}.
Similarly, the results of modifying  $\beta$ and $\gamma$ are shown in Fig.~\ref{fig:figure7} (b) and (c).

\subsection{The balloon force applied to multiple phases}
For physical systems, the fluctuation of data near phase boundaries is maximum. Especially, multiple phase boundaries meet and are more difficult to handle.
Here we discuss whether or not balloon force work in phase diagrams containing more than two phases.

\begin{figure}[tbh]
 \includegraphics[height=3.6cm,width=4.5cm]{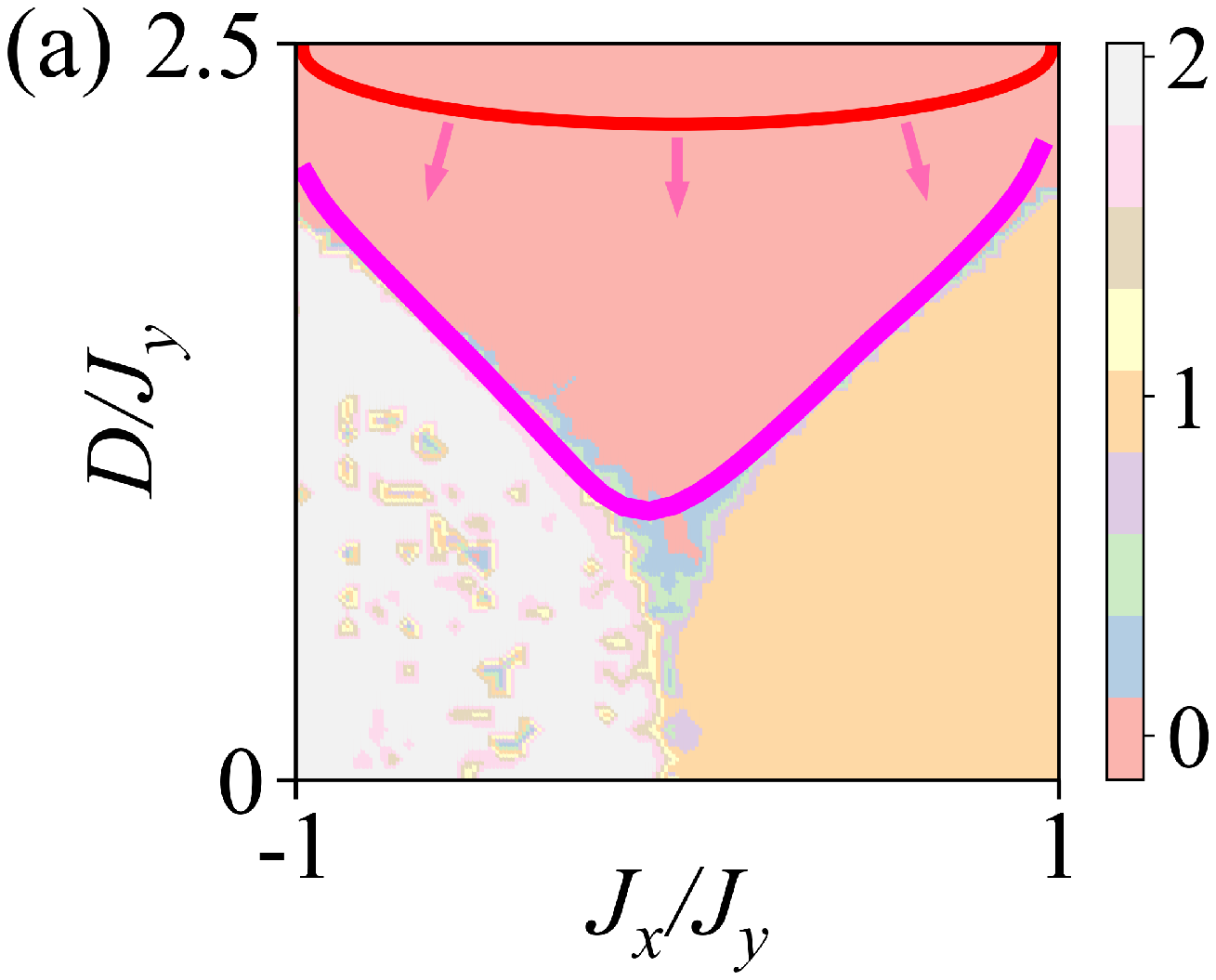}
 \includegraphics[height=3.6cm,width=3.8cm]{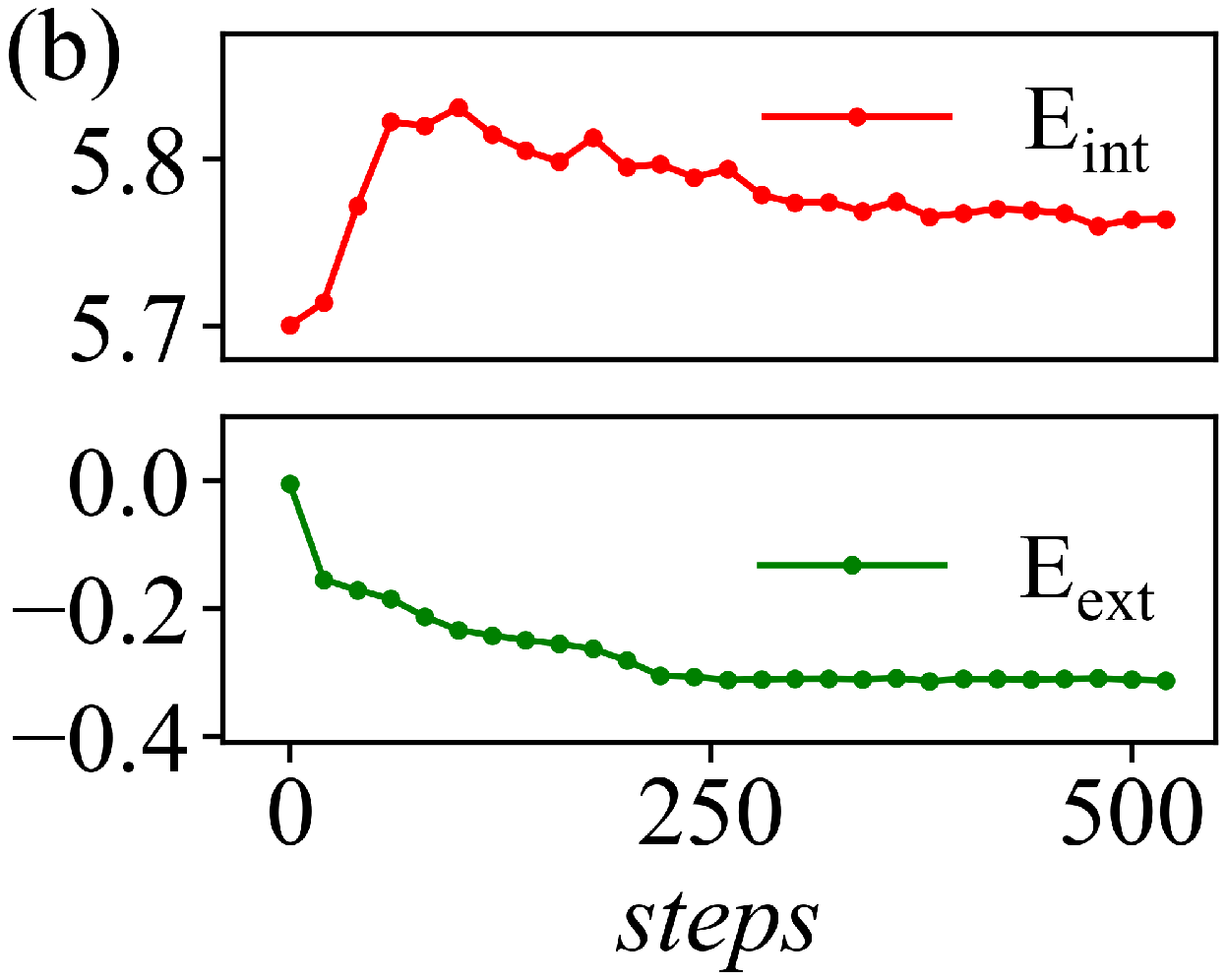}
 \caption{The DCN with balloon force
 is applied to a phase diagram containing three phases.  (a) The initial snake (red), the final snake (pink line), and the balloon force marked by pink arrows. (b) $E_{ext}$ and $E_{int}$ versus steps. }
 \label{fig:figure9}
\end{figure}

In Fig.~\ref{fig:figure9} (a), the balloon force is added to an initial snake that is immersed in the $\begin{smallmatrix} 00\\00 \end{smallmatrix}$ phase and this snake eventually converges to the phase boundary.
Meanwhile, in Fig.~\ref{fig:figure9} (b), $E_{ext}$ and $E_{int}$ are also shown to verify the results.
%We find that the two curves have a peak and a valley respectively.
%The snake is stuck in a  meta-stable position,  which is shown by the red dashed line in Fig.~\ref{fig:figure9} (a).  Eventually, the snake converges to the correct phase boundary.

\begin{figure}[htb] 
  \centering
  \includegraphics[height= 3.4cm,width=8.4cm]{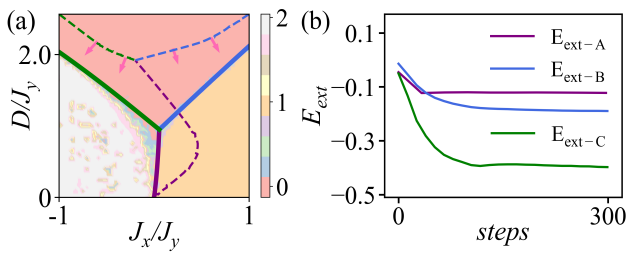} 
  \caption{Introducing balloon force to the SN-DCN. (a) The initial snakes (dashed lines), the final snakes (solid lines), the arrows represent the balloon force. (b) $E_{ext}$ over steps. }
\label{fig:figure13}
\end{figure}

In Fig.~\ref{fig:figure13} (a), the initial snakes are dashed lines, the final snakes are solid lines. Snakes $A$, $B$ and $C$ are marked in purple, blue and green, respectively. By applying balloon force to $snake_{B}$ and $snake_{C}$, respectively, we still get the correct result. In Fig.~\ref{fig:figure13} (b), the $E_{ext}$ converge to a minimum value indicating that the snakes stably stay at the true phase boundary. The results represent that it is feasible to select one or two of the snakes in the SN-DCN to add extra balloon force. 

In short summary, the balloon force can overcome the restriction that the initial position of the snake model must be close to the real boundary; moreover, the balloon force can accelerate the convergence of the snake with the DCN.

%Balloon force helps thtools we developed to be less constrained, and more efficient.

\section{Conclusion and discussion}
\label{sec:con}

In this paper, we extend the DCN with a simple snake model by altering the topology to  a snake net. This potentially
allows one to map out two-dimensional phase diagrams featuring more than two
distinct phases, which is a limitation of the original approach~\cite{liuprl}. Moreover, we introduce an additional external force (balloon force) which may
help the snake to leave its initial position more quickly or leave a wrong initial position and thus may allow for greater freedom in the initialization of the snake.

Unsupervised  machine learning in  studying phase transitions is still active direction~\cite{PhysRevResearch.3.033052,https://doi.org/10.48550/arxiv.2203.06084,gnn}. 
Although the model we tested is one for which the answer is already known, there is a potential value from an unsupervised learning methodological point of view.

It is  meaningful to our method to test {\it higher} dimensional phase diagrams, which are difficult to  search for the boundaries. Related work will be seen elsewhere.
In addition, since the physical systems are limited to  regular lattices,
it is also interesting to apply such a method with a graph neural network~\cite{gnn} for systems on irregular lattices.

{\it Acknowledgement--} 
W. Zhang would like to thank Junyi Xu  for his help and feedback, and contributions of Huijiong Yang and Nan Wu while they were studying in  the office during  their undergraduate years.
This work was supported by the
Hefei National Research Center for Physical Sciences at
the Microscale (KF2021002), and project 12047503  supported by NSFC.  J.Z. is supported by the Natural Science Foundation of
Shanxi Province (Grants No. 202103021224051)

\appendix

\section{Details about the snake model}
\label{sec:appendixA}
In this appendix, we   give a slightly more detailed derivation of the snake model, as well as the iteration matrix for different snakes.

\subsection{Energy Minimization Condition}

In order to get the best position of the  snake, we need to calculate of minimum energy.
Our derivation is slightly different from Ref.~\cite{Butenuth2012}.
We call \(C(s)\) the true contour -- the one we are trying to find. We take some trial contour 
\begin{equation}
  C_{trial}(s) = C(s) + \varepsilon \delta(s)
  \label{eq:contour_not_true}
\end{equation}
which differs from the true contour by \(\varepsilon \delta(s)\)
where \(\varepsilon\) is a small quantity and \(\delta(s)\) is an arbitrary function.
If we write Eq.~\eqref{eq:total_energy} as the following form:
\begin{equation}
	\mathcal{F} = \int_{0}^{1} \La(C(s), C'(s), C"(s)) \dd{s},
  \label{eq:int}
\end{equation}
where \(\La(C, C', C") = E_{int} + E_{ext}\) is the energy per \(\dd s\), $(C(s), C'(s), C"(s))$ 
are considered to be independent variables, then 
we get
\begin{equation}
  \eval{\dv{ \mathcal{F}(\varepsilon)}{\varepsilon}}_{\varepsilon=0} = 0,
  \label{eq:dv}
\end{equation}
since we've already chosen the true contour which makes the total energy minimum
according to the principle of \textit{calculus of variations}.

Putting Eq.~\eqref{eq:int} in Eq.~\eqref{eq:dv},
one gets
  
\begin{widetext}
\begin{align*}
		\eval{\dv{\mathcal{F}(\varepsilon)}{\varepsilon}}_{\varepsilon=0}
		 & =
		\int_{0}^{1}
		\left[
		\pdv{\La}{(C+\varepsilon \delta)} \dv{(C+\varepsilon \delta)}{\varepsilon}
		+
		\pdv{\La}{(C'+\varepsilon \delta')} \dv{(C'+\varepsilon \delta')}{\varepsilon}
		+
		\pdv{\La}{(C''+\varepsilon \delta'')} \dv{(C''+\varepsilon \delta'')}{\varepsilon}
		\right]
		\dd{s} \\
		 & =
		\int_{0}^{1}
		\left[
		\pdv{\La}{C} \delta
		+
		\pdv{\La}{C'} \delta'
		+
		\pdv{\La}{C''} \delta''
		\right]
		\dd{s} \\
		 & =
		 \int_{0}^{1}
		 \pdv{\La}{C} \delta(s)\dd{s}
		 +
		 \pdv{\La}{C'}d\delta(s)
		 +
		 \pdv{\La}{C''}d\delta'(s)
		 \\
		 & =
		 \left.\pdv{\La}{C'}\delta(s)\right|_{0}^{1}
		 +
		 \left.\pdv{\La}{C''}\delta'(s)\right|_{0}^{1}
		 +
		 \int_{0}^{1}
		 \left[
		 \pdv{\La}{C}\delta(s)
		 -
		 \left(\pdv{\La}{C'} \right)'\delta(s)
		 -
		 \left(\pdv{\La}{C''} \right)'\delta'(s)
		 \right]
		 \dd{s}
		 \\
		 & =
		 \int_{0}^{1}
		 \left[
		 \pdv{\La}{C}\delta(s)
		 -
		 \left(\pdv{\La}{C'} \right)'\delta(s)
		 -
		 \left(\pdv{\La}{C''} \right)'\delta'(s)
		 \right]
		 \dd{s}
		 \\
		 & =
		\int_{0}^{1}
		\left[
		\pdv{\La}{C}
		-
		\left(\pdv{\La}{C'} \right)'
		+
		\left(\pdv{\La}{C''} \right)''
		\right]
		\delta(s)
		\dd{s} = 0 
\end{align*}
\end{widetext}

For the arbitrary function \(\delta(s)\), 
the generalized Euler-Lagrange equation can be obtained as:
\begin{equation}
  \pdv{\La}{\vb{C}}
		-
		\left(\pdv{\La}{\vb{C}'} \right)'
		+
		\left(\pdv{\La}{\vb{C}''} \right)''
    =0,
    \label{eq:new_Eulerian_Lagrange_equation}
\end{equation}
 for solving the minimum value of the energy functional.
Eq.~(\ref{eq:new_Eulerian_Lagrange_equation}) is formally a second-order 
Euler-Lagrange equation i.e. a Jacobi-Ostrogradsky formulation \cite{nla.cat-vn43638} 
which has a number of applications in fundamental physics (e.g. Refs.~\cite{PhysRevA.44.1477,doi:10.1139/p99-020,2015PhyEs..28..374S}).

\subsection{Iteration matrix ${\bf A}$}
\label{sec:appendixA2}
By putting the product function Eq.~\eqref{eq:total_energy} into Eq.~\eqref{eq:new_Eulerian_Lagrange_equation}, we get
the differential equation
\begin{equation}
  -  \alpha \vb{C}'' + \beta \vb{C}'''' + \pdv{E_{img}}{\vb{C}} =0.
   \label{eq:one}
\end{equation}
Here we can consider 
\( \alpha \vb{C}'' - \beta \vb{C}''''\)  and -\(\grad E_{img}\)  as 
 internal forces ${\bf F}_{int}$
 and external forces ${\bf F}_{ext}$ on the snake
 respectively.
 The snakes satisfy the mechanical 
 balance ${\bf F}_{int}+{\bf F}_{ext}=0$. 
 %\( \alpha \vb{C}'' - \beta \vb{C}''''\) and external forces -\(\grad E_{img}\) on each node of a snake.
For simplicity, $E_{img} = E_{ext}$ is assumed here, i.e., there is no other external force except the image force. 
Since $E_{img}$ is not available as an expression of $C$ for general images, Eq.~\eqref{eq:one} has no analytical solution. 
Moreover, analytical solutions of higher-order differential
equations are known to generate spurious or unstable solutions as evidenced
by the Ostrogradsky instability. However, this equation can be reliably solved by the finite difference numerical method and reads, 
\begin{align}
     	&\frac{\partial E_{img}}{\partial C}+\alpha((C_{i}-C_{i-1})-(C_{i+1}-C_{i}))\nonumber\\
     	+&\beta(C_{i-2}-2C_{i-1}+C_{i})-2\beta(C_{i-1}-2C_{i}+C_{i+1})\nonumber\\
     	+&\beta(C_{i}-2C_{i+1}+C_{i+2})=0.
    \label{eq:EQA6}
\end{align}
Eq.~\eqref{eq:EQA6} is the mechanical equation satisfied by the 
node $i$ and its neighborhood nodes. By combining the set of equations for all nodes together, the following equation can be obtained,
     \begin{equation}
     	{\bf A}C+\frac{\partial E_{img}}{\partial C}=0,
     	\label{eq:EQA7}
     \end{equation} 
where ${\bf A}$ is a pentadiagonal banded matrix, which only depends on the parameters $\alpha$ and $\beta$, and $C=[C_0, C_1, \cdots, C_{N-1}]^{Transpose}$.
For a snake with {\it periodic} boundary conditions,
the matrix is defined as:

\begin{align}
\bf A=\mqty[
c_0   & b_1   & a_2 &         &         &         & a_{N-2} & b_{N-1} \\
b_0   & c_1   & b_2 & a_3     &         &         &         & a_{N-1} \\
a_0   & b_1   & c_2 & b_3     & a_4     &         &         &         \\
      & a_1   & b_2 & c_3     & b_4     & a_5     &         &         \\
      &       &     &         & ...     &         &         &         \\
      &       &     & a_{N-5} & b_{N-4} & c_{N-3} & b_{N-2} & a_{N-1} \\
a_{0} &       &     &         & a_{N-4} & b_{N-3} & c_{N-2} & b_{N-1} \\
b_{0} & a_{1} &     &         &         & a_{N-3} & b_{N-2} & c_{N-1}],
\label{eq:simple}
\end{align}
where the values of \(a_i\),~ \(b_i\),~ \(c_i\) ~are as follows:
\begin{subequations}
	\begin{align}
	a_{i}&=\beta, \\
	b_i&=-\alpha-4\beta, \\
	c_i&=2\alpha+6\beta .
	\end{align}
\end{subequations}
Pentadiagonal matrices are sparse band matrices and therefore useful for numerical analysis.

For the simple SN model, such as $C^A$ with one fixed node at the end and one common node at the other end as shown in Fig.~\ref{fig:figure10} (a), the  matrix ${\bf A}$ 
is modified as
\begin{align}
\bf A=\mqty[
      &       &     &         &         &         &         &         \\
d_0   & e_1   & b_2 & a_3     &         &         &         & a_{N-1} \\
a_0   & b_1   & c_2 & b_3     & a_4     &         &         &         \\
      & a_1   & b_2 & c_3     & b_4     & a_5     &         &         \\
      &       &     &         & ...     &         &         &         \\
      &       &     & a_{N-5} & b_{N-4} & c_{N-3} & b_{N-2} & a_{N-1} \\
a_0   &       &     &         & a_{N-4} & b_{N-3} & e_{N-2} & d_{N-1} \\
      &       &     &         &         & f_{N-3} & g_{N-2} & f_{N-1}],
      \label{eq:sn-dcn}
\end{align}
where the elements of the first row are 0 due to the first node being fixed.
Unlike the periodic boundary snake, there are many boundary related elements that have been revised and  the values of \(a_i\),~\(b_i\),~\(c_i\),~\(d_i\),~\(e_i\),~\(f_i\),~\(g_i\)~are as follows:
\begin{subequations}
	\begin{align}
	a_i&=\beta,                      \\
	b_i&=-\alpha-4\beta,             \\
	c_i&=2\alpha+6\beta,             \\
	d_i&=-\alpha-2\beta,                     \\
	e_i&=2\alpha+5\beta,                   \\
	f_i&=\xi,                    \\
	g_i&=-2\xi.                 
	\end{align}
\end{subequations}
Similarly,  for the snake which has two movable endpoints such as $C^C$ as shown in Fig.~\ref{fig:figure11} (a), the matrix is:
\begin{align}
\bf A=\mqty[
f_0   & g_1   & f_2 &         &         &         &         &         \\
d_0   & e_1   & b_2 & a_3     &         &         &         & a_{N-1} \\
a_0   & b_1   & c_2 & b_3     & a_4     &         &         &         \\
      & a_1   & b_2 & c_3     & b_4     & a_5     &         &         \\
      &       &     &         & ...     &         &         &         \\
      &       &     & a_{N-5} & b_{N-4} & c_{N-3} & b_{N-2} & a_{N-1} \\
a_0   &       &     &         & a_{N-4} & b_{N-3} & e_{N-2} & d_{N-1} \\
      &       &     &         &         & f_{N-3} & g_{N-2} & f_{N-1}],
      \label{eq:ed-sn-dcn}
\end{align}
Ref.~\cite{
Butenuth2012} defines a total matrix ${\bf A}$ which contains the elements for all snakes.
Here the matrix ${\bf A}$ we defined, is for each snake. 
The coupling between each snake can be achieved by passing the coordinates of common nodes from one snake to other snakes.

\subsection{The iteration equation}
\label{sec:appendixA3} 
Eq.~\eqref{eq:EQA7} is a static mechanical equilibrium equation without considering the damping force of the deformed contour. To  describe a dynamic contour,  the time parameter $t$ and the damping force \(F_{dump}(C_t) =-\gamma\pdv{C_t}{t}\) 
and the inertia term $\mu\pdv[2]{C_t}{t}$ are introduced, then the following equation,
\begin{equation}
    \mu\pdv[2]{C_t}{t}=F_{dump}(C_t))+F_{int}(C_t)+F_{img}(C_t),
\end{equation}
is obtained, where
\begin{subequations}
	\begin{align}
  -F_{int}&=-\alpha C"_t+\beta C""_t={\bf A}C,\\
  -F_{img}&=  \frac{ \partial E_{img}}{\partial C} = \eta f(C_t),
  %F_{dump}&=-\gamma\pdv{C(s,t)}{t} .
	\end{align}
\end{subequations}
and where $\eta$ is an additional parameter in order to control the weight between internal and image energy.
The inertia term $\mu\pdv[2]{C_t}{t}$ is set to zero because the inertia term  can 
 cause the snake to cross the target boundary.   This dynamic process 
becomes quasi-static 
process as:
\begin{equation}
    \gamma\pdv{C_t}{t}=F_{int}+F_{img}.
    \label{eq:EQA15}
\end{equation} 
The equation of the discrete snake becomes:
\begin{equation}
   - \frac{\gamma(C_{t+1}-C_{t})}{\Delta t}={\bf A}C_{t+1} + \eta f(C_t).
\end{equation}
Taking the time step as  $\Delta t= 1$, one gets
\begin{equation}
    -\gamma(C_{t+1}-C_{t})={\bf A} C_{t+1} +\eta f(C_{t})
    \label{eq:EQA17}.
\end{equation}  
Finally, the iteration equation of the snake is obtained as:
\begin{equation}
     	C_{t+1}=({\bf A}+\gamma {\bf I})^{-1}[\gamma C_{t}-\eta f(C_{t})]
     	\label{eq:EQA18}, 
     \end{equation} 
where $I$ is the identity matrix.

\subsection{Segmentation of images}

\begin{figure}[tb]
 \includegraphics[height=3.8cm,width=5.8cm]{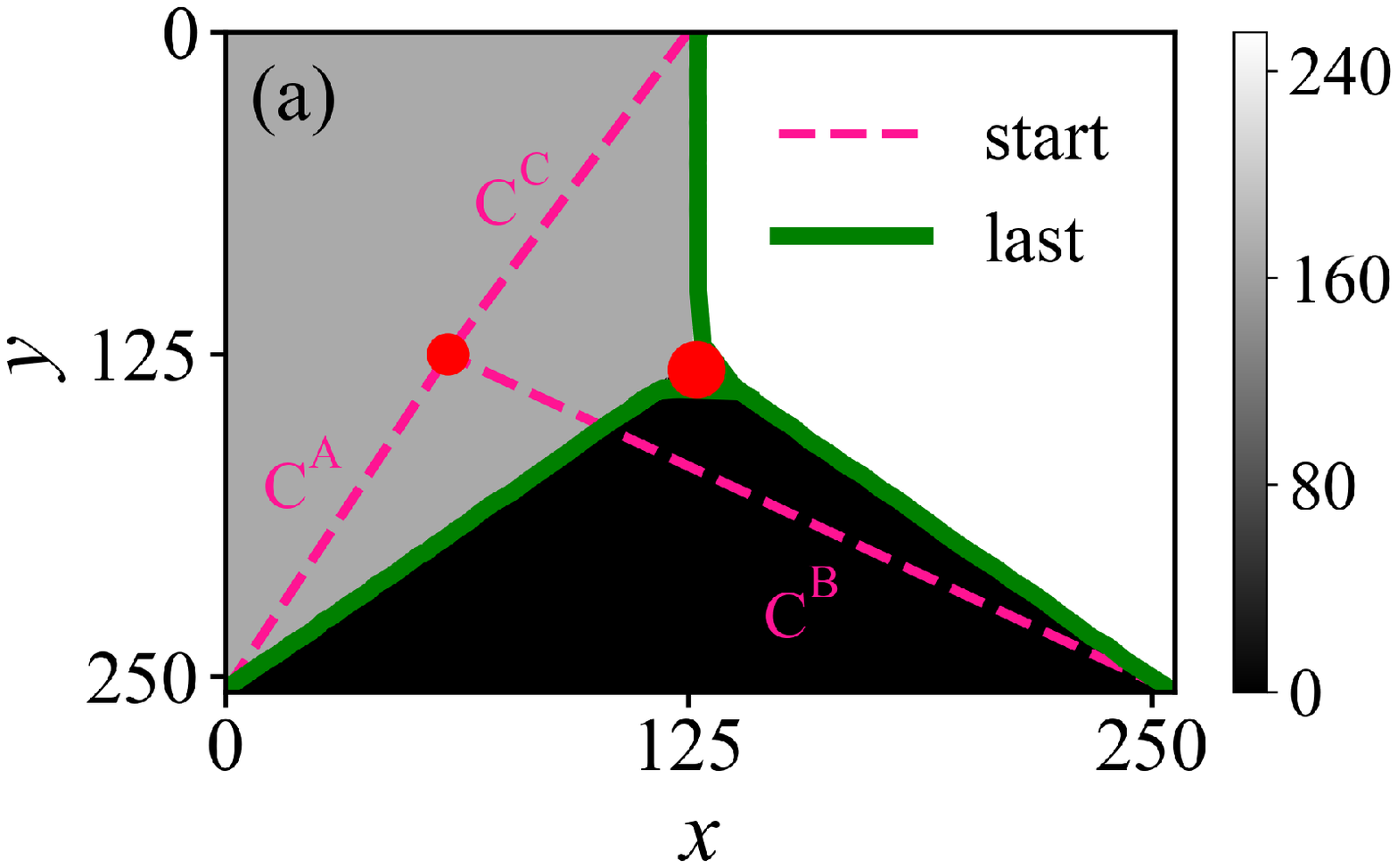}
 \vskip -0.13 cm
 \includegraphics[height=3.8cm,width=8.2cm]{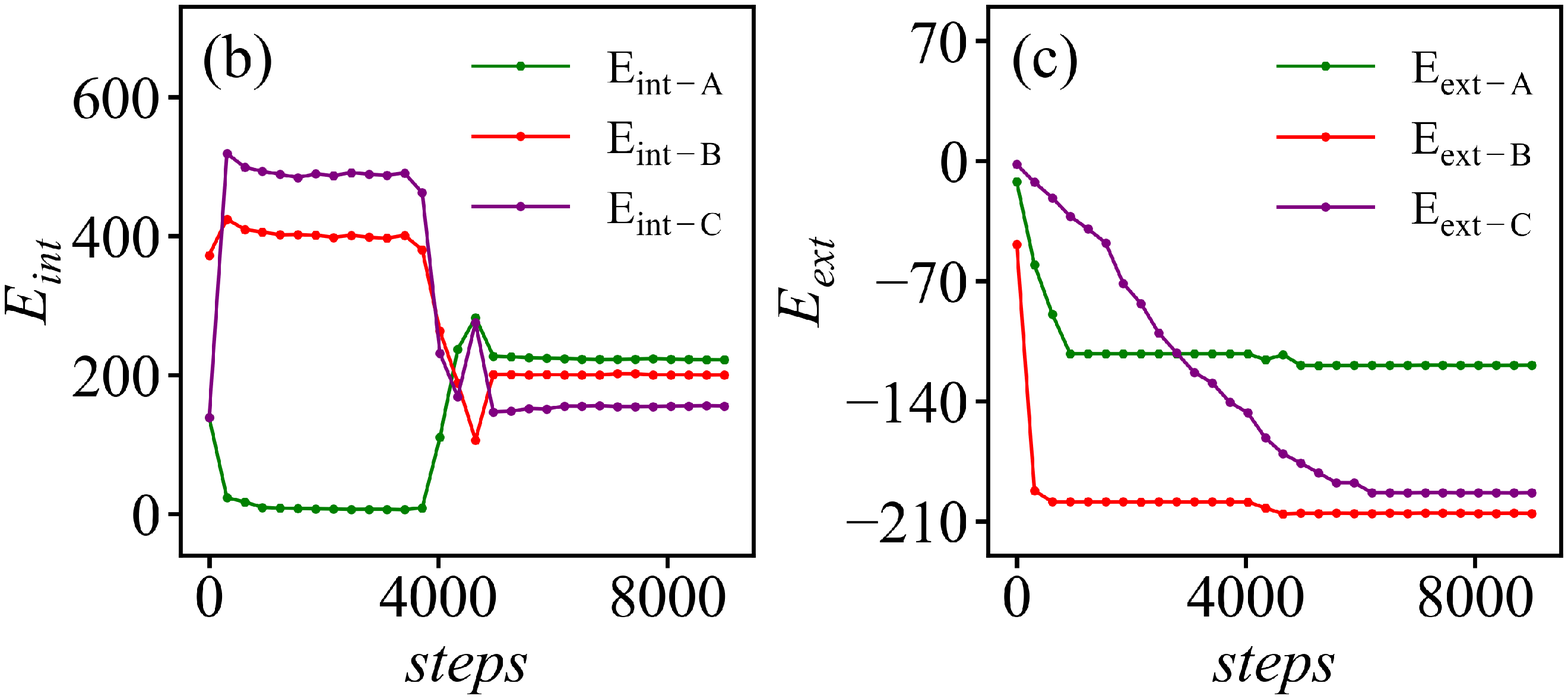} 
 \caption{ The results of applying the SN  without the DCN to detect contour in image with three different colors. (a) The gray image, the initial snakes
 are in pink, and the final snakes are in green. The red dot represent the common nodes  where the three snakes intersect. (b) $E_{int}^{A}$ - $E_{int}^{C}$ versus steps. (c)   $E_{ext}^{A}$ - $E_{ext}^{C}$ versus steps.}
\label{fig:figure10}
\end{figure}

\begin{figure}[tb]
  \includegraphics[height=3.7cm,width=6.2cm]{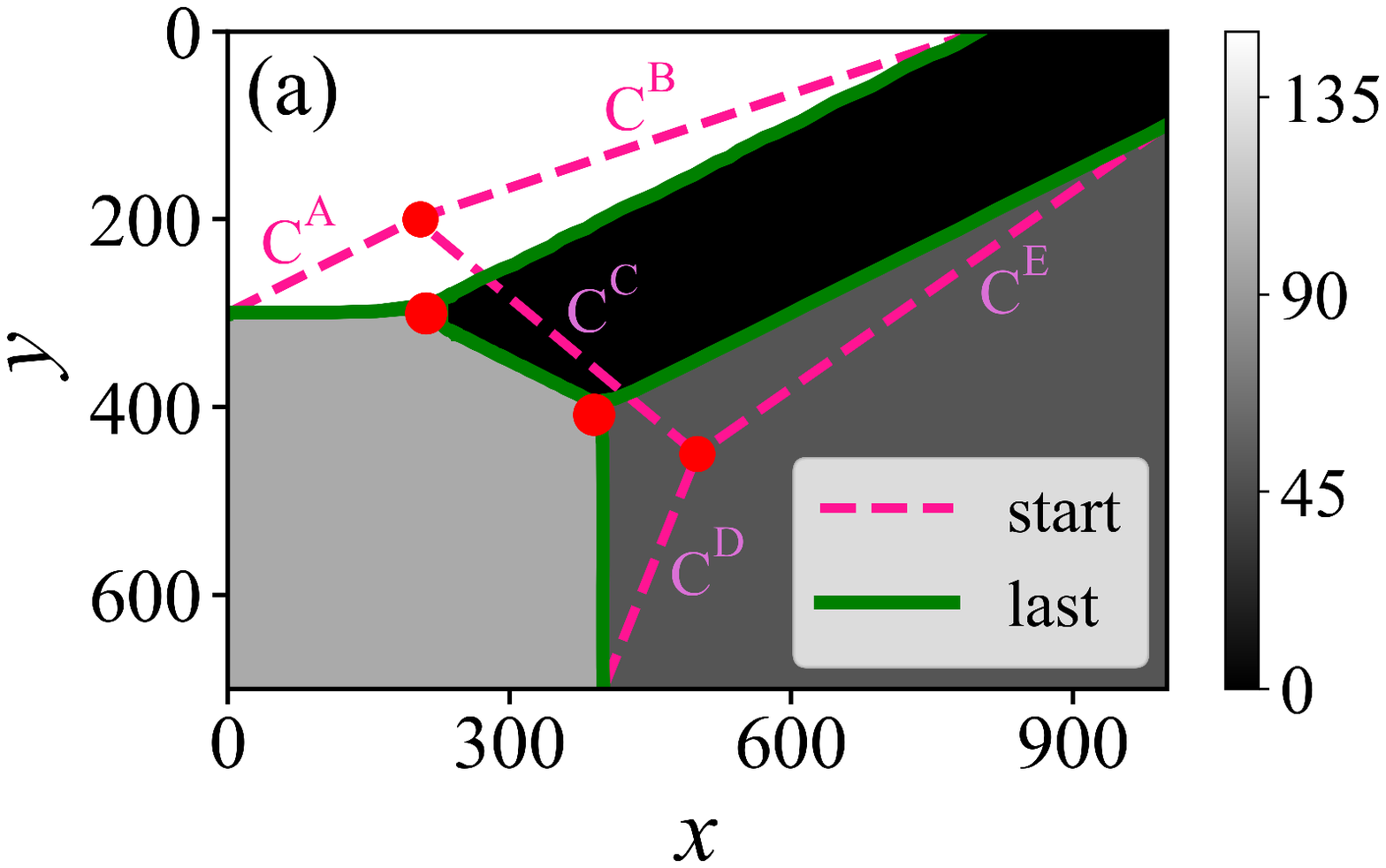}
  \vskip -0.13 cm
  \includegraphics[height=4.cm,width=4.2cm]{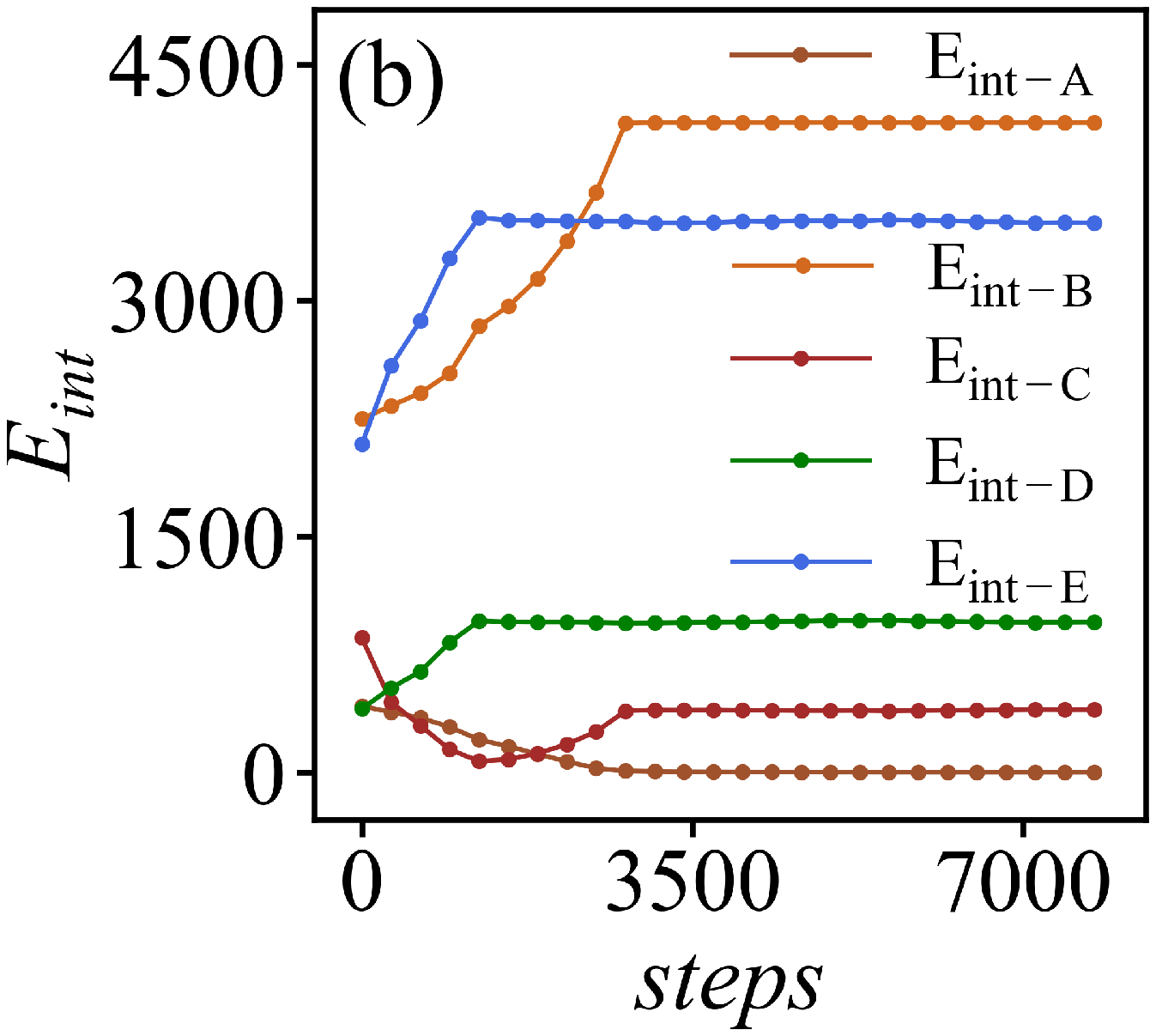} 
  \hspace{-0.17 cm}
  \includegraphics[height=4.cm,width=4.2cm]{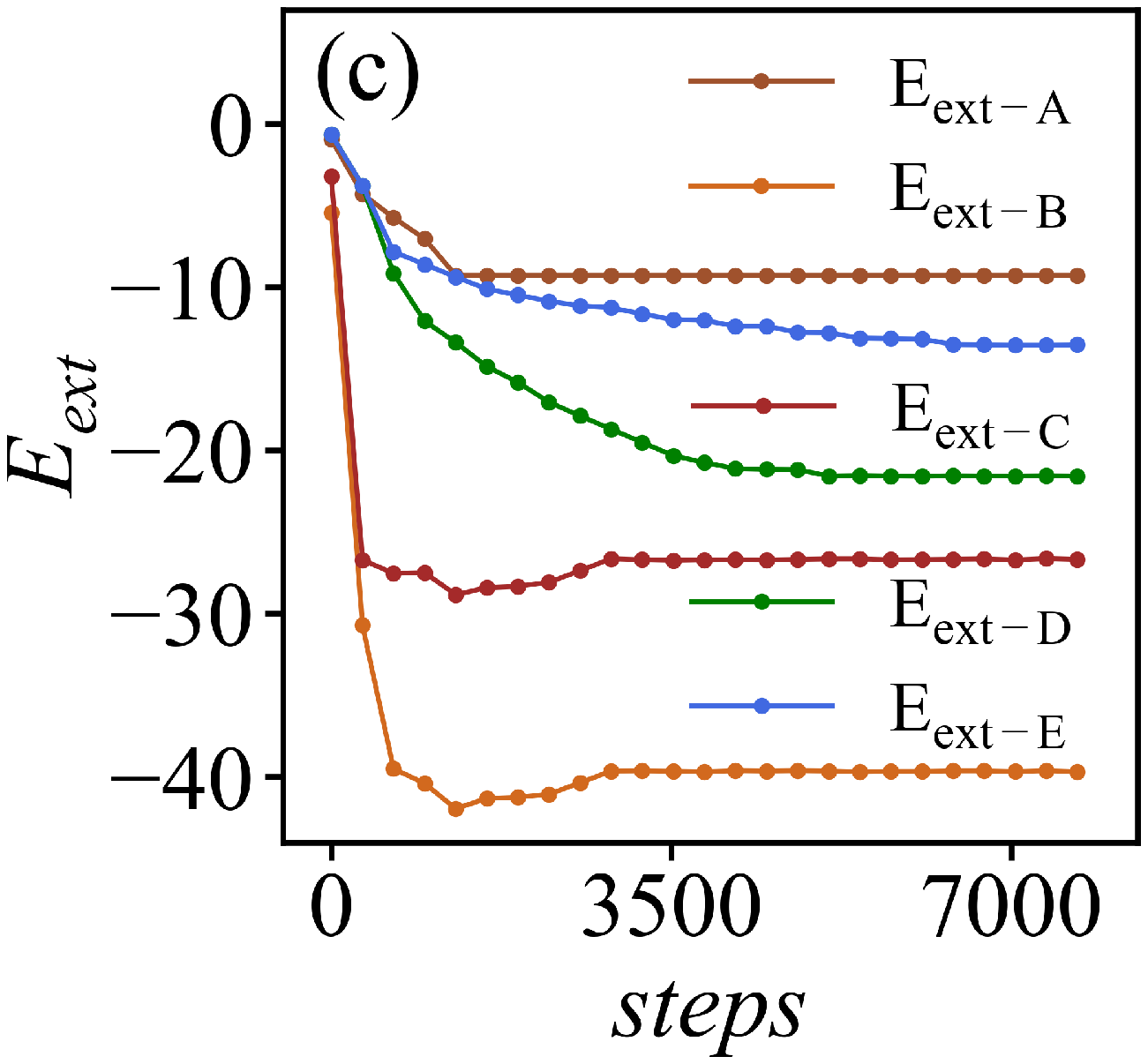}
 \caption{ The results of applying the SN  without the DCN to detect contour in image with four different colors. (a) The gray image, initial snakes (pink), final snakes (green) and the common nodes (red dots), (b) $E_{int}^A$ - $E_{int}^E$ versus steps, (c)   $E_{ext}^A$ - $E_{ext}^E$ versus steps.}
\label{fig:figure11}
\end{figure}

In  Fig.~\ref{fig:figure10} (a), a gray image with three different values of the pixels are shown as white, gray and black. 
The dashed lines are the initial snakes and 
the solid lines are final snakes.
It is clear that, for the pure image, 
the SN model can reach the boundaries between the different color blocks.

In Fig.~\ref{fig:figure10} (b) and (c),
the external and internal energies of each snake have been given with the number of iterative steps. All the quantities converge very well. The internal energy converges and this means the  shapes of the snakes are no longer changing and the external energy converges to a minimum value implying that the snakes move to the boundary to be found. The stabilization of both internal and external energies indicates that the snake stays steadily at the boundary to be sought. In Fig.~\ref{fig:figure11} (a)-(c), similar results for more a general SN are shown. 

Although this task of contour extraction belongs to the field of computer vision and image processing, it is helpful to understand the SN-DCN.
 
\section{Parameters of the neural networks}
\subsection{The parameters for obtaining Fig.~\ref{fig:figure3}}
\label{sec:par1}
The parameters of the neural network $\mathcal{N}$ are set as follows:  mini-batch size $N_{b}=1500$, initial learning rate $\alpha_{\mathcal{N}}=0.01$, learning rate decay = 0.999, input layer 256, hidden layer 160, output layer 2, 
and optimizer='ADAM'.  Here the ``Adam'' optimizer is implemented with  the TensorFlow library~\cite{tensorflow2015-whitepaper}. The parameters $\alpha$, $\beta$, $\gamma$ of $snake_{A}- snake_{C}$, are set to $\alpha=0.05, 0.2, 0.2$ $\beta=10, 5, 5$, and $\gamma=0.6, 0.1, 0.1$, respectively. $\xi=0.1$. The dynamic unit width $\sigma$ of each node is initialized to $0.05$ and is limited in the range from $0.07$ to $0.01$.

\subsection{The parameters for obtaining Fig.~\ref{fig:figure4}}
\label{sec:par2}
The parameters of the neural network $\mathcal{N}$ are set as follows:  mini-batch size $N_{b}=1500$, initial learning rate $\alpha_{\mathcal{N}}=0.01$, learning rate decay = 0.997, input layer 256, hidden layer 160, output layer 2, %dropout keep probability 0.8, 
and optimizer='ADAM'. The parameters $\alpha$, $\beta$, $\gamma$ of $snake_{A}$, were set to $\alpha=5$, $\beta=5$, and $\gamma=0.2$. The parameters of $snake_{B}- snake_{E}$, are set to $\alpha=2$, $\beta=0.4$, and $\gamma=0.22$. $\xi=0.2$. The dynamic unit width $\sigma$ of each node is initialized to $0.06$ and is limited in the range from $0.08$ to $0.02$.

\bibliography{ref}
\end{document}